\documentclass[conference]{IEEEtran}
\IEEEoverridecommandlockouts
\usepackage{cite}
\usepackage{amsmath,amssymb,amsfonts}
\usepackage{algorithmic}
\usepackage{graphicx}
\usepackage{textcomp}
\usepackage{xcolor}

\usepackage{soul}
\usepackage[utf8]{inputenc}
\usepackage{multirow}
\usepackage{booktabs}
\usepackage{epstopdf}
\usepackage{url}
\usepackage {footnote} 
\usepackage{fancyhdr}
\makesavenoteenv {tabular}
\graphicspath{{figures/}}

\newcommand{\mengdi}[1]{\textcolor{black}{#1}}

\def\BibTeX{{\rm B\kern-.05em{\sc i\kern-.025em b}\kern-.08em
    T\kern-.1667em\lower.7ex\hbox{E}\kern-.125emX}}
\begin{document}


\title{Characterizing Deep Learning Training Workloads on Alibaba-PAI}


\author{
Mengdi Wang$^1$, 
Chen Meng$^1$, 
Guoping Long$^1$,
Chuan Wu$^2$,
Jun Yang$^1$,
Wei Lin$^1$,
Yangqing Jia$^1$
\\ 
$^1$ Alibaba Group,
$^2$ The University of Hong Kong\\
\{didou.wmd, mc119496, guopinglong.lgp\}@alibaba-inc.com, \\
cwu@cs.hku.hk,
\{muzhuo.yj, weilin.lw, yangqing.jia\}@alibaba-inc.com
}

\maketitle
\pagestyle{empty}

\begin{abstract}
Modern deep learning models have been exploited in various domains, including computer vision (CV), natural language processing (NLP), search and recommendation.
In practical AI clusters, workloads training these models are run using software frameworks such
as TensorFlow, Caffe, PyTorch and CNTK. One critical issue for efficiently operating practical AI clouds, is to characterize the computing and data transfer demands of these workloads, and more importantly, the training performance 
given the underlying software framework and hardware configurations.
In this paper, we characterize deep learning training workloads from 
Platform of Artificial Intelligence (PAI) in Alibaba.
We establish an analytical framework to investigate detailed execution time breakdown of various workloads using different training architectures, to identify performance bottleneck. Results show that weight/gradient communication during training takes almost 62\% of the total execution time among all our workloads on average.
The computation part, involving both GPU computing and memory access,
are not the biggest bottleneck based on collective behavior of the workloads.
We further evaluate attainable performance of the workloads on various potential software/hardware mappings, and explore implications on software architecture selection and hardware configurations.
We identify that 60\% of \emph{PS/Worker} workloads can be potentially sped up when ported to the \emph{AllReduce} architecture exploiting the high-speed NVLink for GPU interconnect, and on average 1.7X speedup can be achieved when Ethernet bandwidth is upgraded from 25 Gbps to 100 Gbps.
\end{abstract}

\section{Introduction}
Recent years have witnessed the proliferation of deep learning models used in various domains of the industry, including image processing \cite{krizhevsky2012imagenet,he2016identity},
video understanding \cite{karpathy2014large,simonyan2014two}, language understanding \cite{bahdanau2014neural,devlin2018bert}, speech recognition \cite{graves2013speech,chorowski2015attention}, commodity
search and recommendation \cite{wang2018billion,ying2018graph}, autonomous drive \cite{chen2015deepdriving}, and various others \cite{silver2016mastering,zoph2016neural}. Large IT companies are investing substantially to build large AI clouds/clusters, equipped with expensive hardware such as GPUs, to run various deep learning workloads to support their AI-driven services.

This paper presents a characterization of the workloads from Platform of Artificial Intelligence (PAI) in Alibaba.
PAI is a ML(machine learning)-as-a-service platform that simplifies machine learning adoption and makes large-scale AI to meet the needs of Alibaba internel business. It has also been shipped to Aliyun as a cloud product to serve public users.
Thousands 
of training jobs are submitted to PAI on a daily basis, with different business objectives, and diversified computing, communication and I/O requirements and constraints. This paper focuses on one critical aspect of these practical workloads: characterize various resource requirements and identify performance bottlenecks given software frameworks and hardware configurations. The observations are intended to instruct exploration of the workload
optimization space, and guide software and hardware configurations/provisioning, 
to improve workload execution performance. 

Existing AI workload characterization work mostly focus on quantitative, precise performance modeling of AI workloads \cite{shi2016benchmarking,gu2017deepprof} or  building benchmark platforms to measure model performance \cite{adolf2016fathom,li2018tartan,gao2018data2} (see Sec.~\ref{sec_related} for detailed discussions).
We take a different angle, collectively characterizing behavior of thousands of training jobs in a production cluster, as well as projecting potential performance gains with different software architectures and hardware configurations based on a simple analytical model.
Contributions of this work are summarized as follows:

\textbf{First}, we 
present a lightweight framework to characterize the production workloads at the  cluster level. 
We comprehensively include 
 not only the basic aspects of computation and weight communication in training jobs, as considered in previous studies \cite{hazelwood2018applied,adolf2016fathom}, but also the input data I/O aspects. 
 Our analysis shows that the data I/O time is non-negligible, 
 especially for single-node training workloads; for distributed workloads, input data I/O can potentially become the performance bottleneck 
 after gradient communication has been optimized.

\textbf{Second}, our statistical analysis of the cluster workloads reveals that multi-GPU interconnect rather than the computation power is more likely the bottleneck under the current widely adopted training architectures and system configurations.
Previous work largely focus on analyzing computation resource and memory access of AI workloads \cite{park2018deep,adolf2016fathom}.
Shi \emph{et al.}~\cite{shi2018performance} study the communication factor,
and make a similar conclusion that the current DL frameworks,
including TensorFlow, CNTK and MXNet, do not perform well in scalability via Ethernet interconnect; their analysis is mainly focusing on 
performance comparison among different DL frameworks.
Instead, we investigate the impact of data traffic on workload performance by collectively investigating a large number of training jobs, and explore potential optimization approaches for communication reduction.


\textbf{Third}, we establish simple analytical performance models based on the key workload features, aiming at 
exposing fundamental performance bottlenecks. 
Our analytical modeling is different from previous characterization methods \cite{park2018deep,zhu2018benchmarking,adolf2016fathom},
most of which adopt actual runtime profiling measurements 
for bottleneck analysis. Based on the analytical models, we estimate potential performance gains if the workloads were running on different software architectures and hardware configurations. The focus is to investigate which system architecture (\emph{PS/worker} or \emph{AllReduce}) should be adopted, how much benefits high-speed multi-GPU interconnect, NVLink, may bring, and how performance bottlenecks may shift with different architecture and hardware configurations. 

\textbf{Finally}, we conduct detailed analysis of representative deep learning workloads using both analytical models and testbed experiments, 
in the domains of commodity embedding, search and recommendation, \emph{etc}. 
The relevant models are becoming more and more important in companies related to e-commerce, social networking and search engines, and in PAI, consume a large fraction of resources. Results of the case studies show that differences between estimated performance using our analytical method and actual measurements
are less than 10\% on average.
\mengdi{Based the basic workload features, we explore different optimization techniques upon different types of workloads, including mixed-precision training with TensorCore \cite{volta}, operation fusion via XLA \cite{XLA} and also changing the system architectures.}
We summarize useful observations and implications on improving practical deep learning training workloads.

\section{Background and Methodology}
\label{sec_methodology}
We first present our workload characterization framework. 
While the characterization framework is established based on TensorFlow \cite{abadi2016tensorflow}, the methodology
applies to other frameworks \cite{jia2014caffe,paszke2017automatic,chen2015mxnet,yu2014introduction} as well.

\subsection{Architecture Components Modeling}
\label{sec_method_background}
\vspace{-0.2cm}
\subsubsection{System Infrastructure \& Configuration}
\label{sec_system_infra}

\begin{figure}[!htb]
\vspace{-0.3cm}
\centering
	\begin{minipage}[b]{0.8\linewidth}
	\centerline{\includegraphics[width=\linewidth]{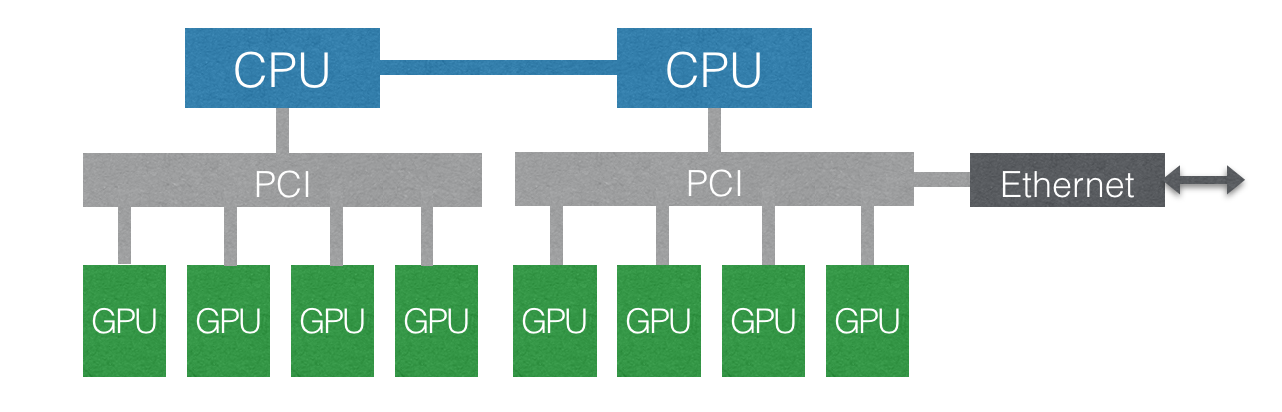}}
	\centerline{\scriptsize (a) Server without NVLink}
	\end{minipage}
	\vspace{0.2cm}
	\begin{minipage}[b]{0.8\linewidth}
	\centerline{\includegraphics[width=\linewidth]{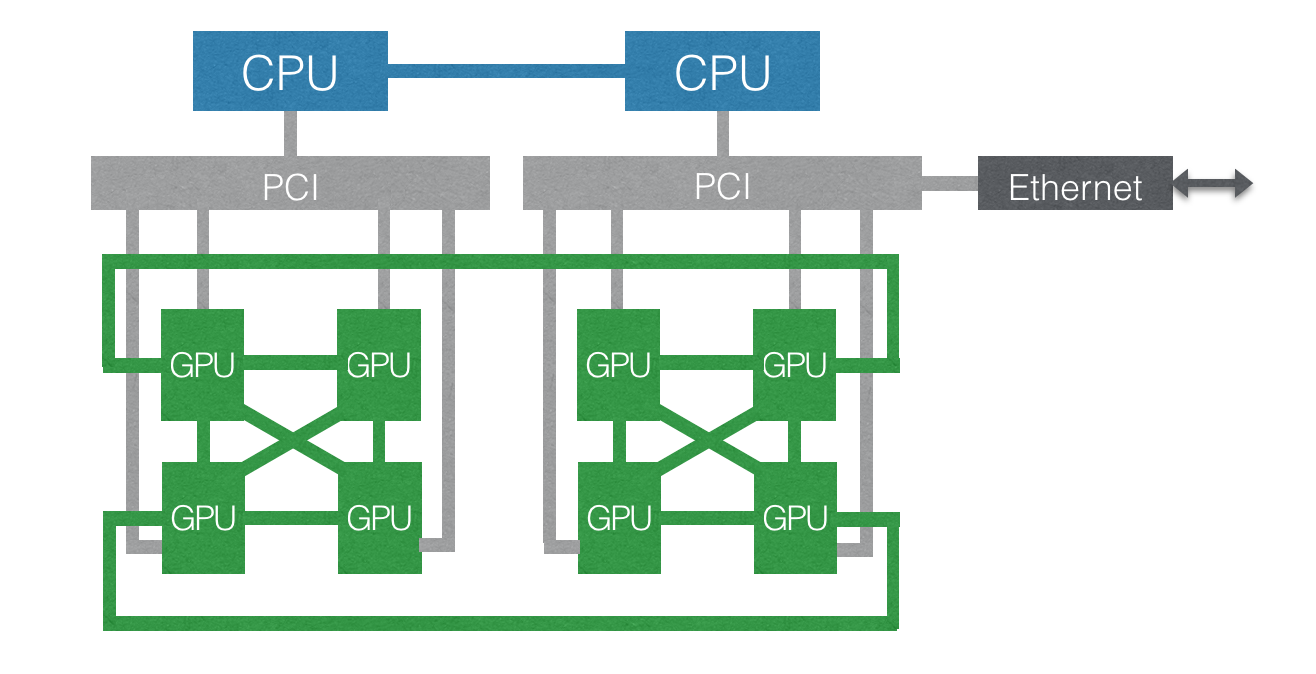}}
	\centerline{\scriptsize (b) Server with NVLink}
	\end{minipage}
	\vspace{-0.2cm}
	\caption{System Infrastructure.}
	\label{PAI_infrastructure}
	\vspace{-0.2cm}
\end{figure}

Figure \ref{PAI_infrastructure} shows the basic server configurations in the AI cluster.  There are typically two types of multi-GPU servers, equipped with/without NVLink \cite{NVLink}. 
The NVLink technology provides high-speed interconnect across multiple GPUs with a `hybrid mesh grid' topology, 
as show in Fig. \ref{PAI_infrastructure}(b), to resolve the bandwidth bottleneck of PCIe interconnect among the GPUs. 
Due to cost issue, servers in some sub-clusters of PAI are equipped with NVLink, while others are not yet.

The basic server configuration where we collect the workload traces is shown in Table \ref{table_baseline_config}. 
The servers are interconnected via bi-directional 25Gbps Ethernet. We will further discuss the impact of the system configurations through varying the configuration settings in Sec.~\ref{subsec_opt_space}.

\begin{table}[!htbp]
\vspace{-0.2cm}
\caption{System Settings.}
\label{table_baseline_config}
\vspace{-0.2cm}
\centering
\begin{tabular}{ c| c |c }
\hline

 \multirow{2}{*}{GPU}& FLOPs & 11 TFLOPs  \\
 \cline{2-3}
 & Memory & 1 TB / second\\
 \hline
 \multirow{3}{*}{Bandwidth} 
 &Ethernet & 25 Gb / second\\
 \cline{2-3}
 & PCI & 10 GB / second\\
  \cline{2-3}
 & NVLink & 50 GB / second\\
 \hline
\end{tabular}
\vspace{-0.2cm}
\end{table}

\subsubsection{System Architecture}
\label{sec_parallel}
More than 85\% computation resources on our cluster are used by distributed training workloads.
DL training workloads can be parallelized via data parallelism, model parallelism and also hybrid parallelism \cite{mayer2019scalable}. 
While model parallelism and hybrid parallelism enable training neural networks which a single processor cannot support, they usually require significant human efforts for efficient model partition.
Data parallelism is more model agnostic, 
and has been the most widely used paradigm for parallelizing neural network training \cite{shallue2018measuring}.
We focus on data-parallel training jobs in this work.

\begin{figure}[!htb]
\vspace{-0.3cm}
\centering
	\begin{minipage}[b]{0.7\linewidth}
	\centerline{\includegraphics[width=\linewidth]{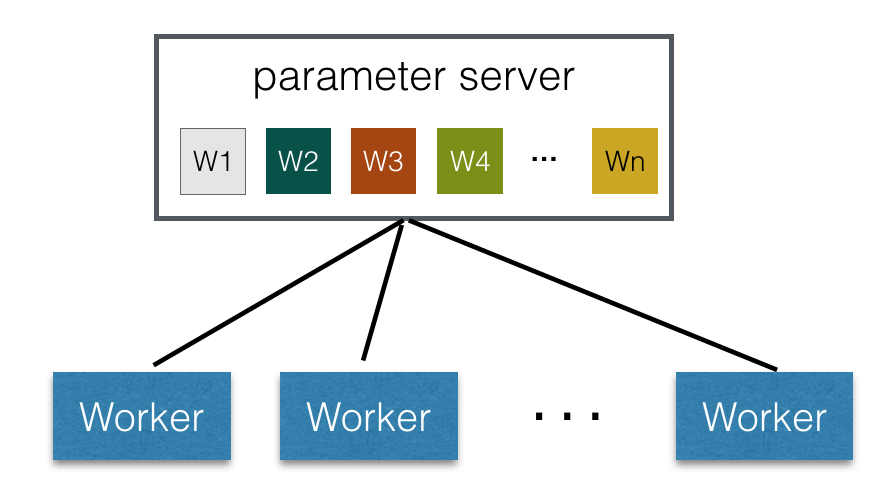}}
	\centerline{\scriptsize (a) \emph{PS/Worker}}
	\end{minipage}
	\begin{minipage}[b]{0.7\linewidth}
	\centerline{\includegraphics[width=\linewidth]{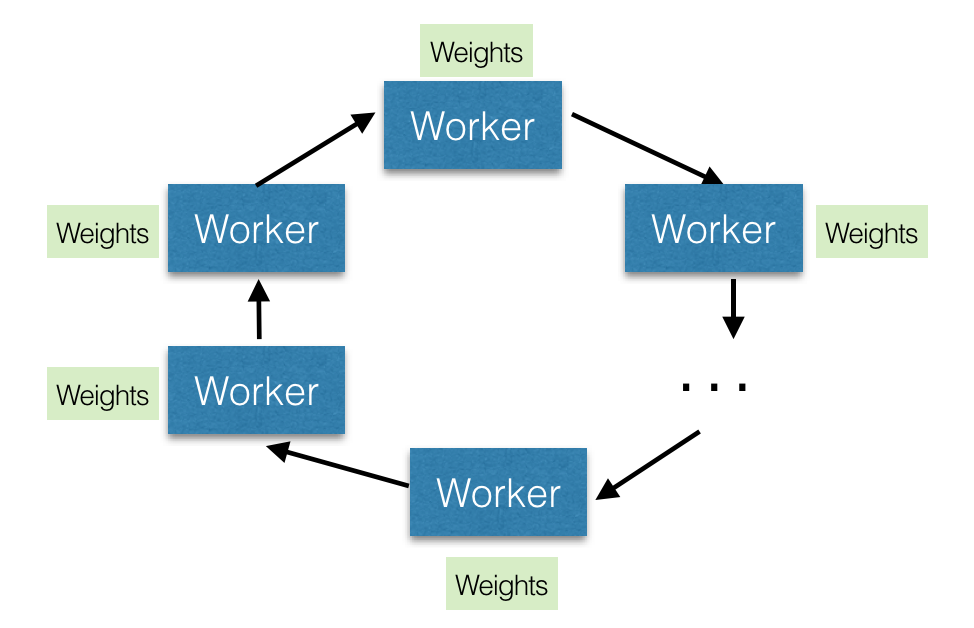}}
	\centerline{\scriptsize (b) \emph{AllReduce}}
	\end{minipage}
	\vspace{-0.2cm}
	\caption{System Architecture.}
	\label{fig_PAI_parallel}
	\vspace{-0.2cm}
\end{figure}

There are two types of system architectures, centralized and decentralized, for synchronizing weights/gradients among distributed training replicas.
In a (parameter) centralized architecture, represented by the parameter server (PS) architecture \cite{li2014scaling}, one or multiple parameter servers manage the gradient aggregation, and each worker holds a training replica,
pulling variables from the PSs at the beginning in each training step
and pushing gradients back to them at the end of each step.
In a (parameter) decentralized architecture, 
the global parameters are placed replicated or partitioned on all training nodes;  
each training node exchanges the gradients via an \emph{AllReduce} operation at the end of each training iteration.
This architecture can benefit from the NVIDIA Collective Communications Library (NCCL) \cite{nccl2018} for high-speed multi-node/multi-GPU communication. In this paper, we have implemented a new decentralized parallel training strategy called PEARL to handle large embedding weights. The detailed discussions about PEARL are showed in Sec. \ref{sec_pearl_arch}.

Currently, representative deep learning frameworks such as TensorFlow, Pytorch and MXNet mainly support the decentralized architecture in the replica mode: all model parameters are replicated to each device and data parallelism is used with the AllReduce algorithnm.
In our cluster, roughly 29\% jobs are running using the PS architecture and less than 1\% using AllReduce, as we adopt AllReduce only after our cluster are equipped with NVLink. 

\subsubsection{DL Training Workloads}
\label{sec_workload}

\begin{figure}[!htb]
\vspace{-0.3cm}
\centering
	\includegraphics[width=0.9\linewidth]{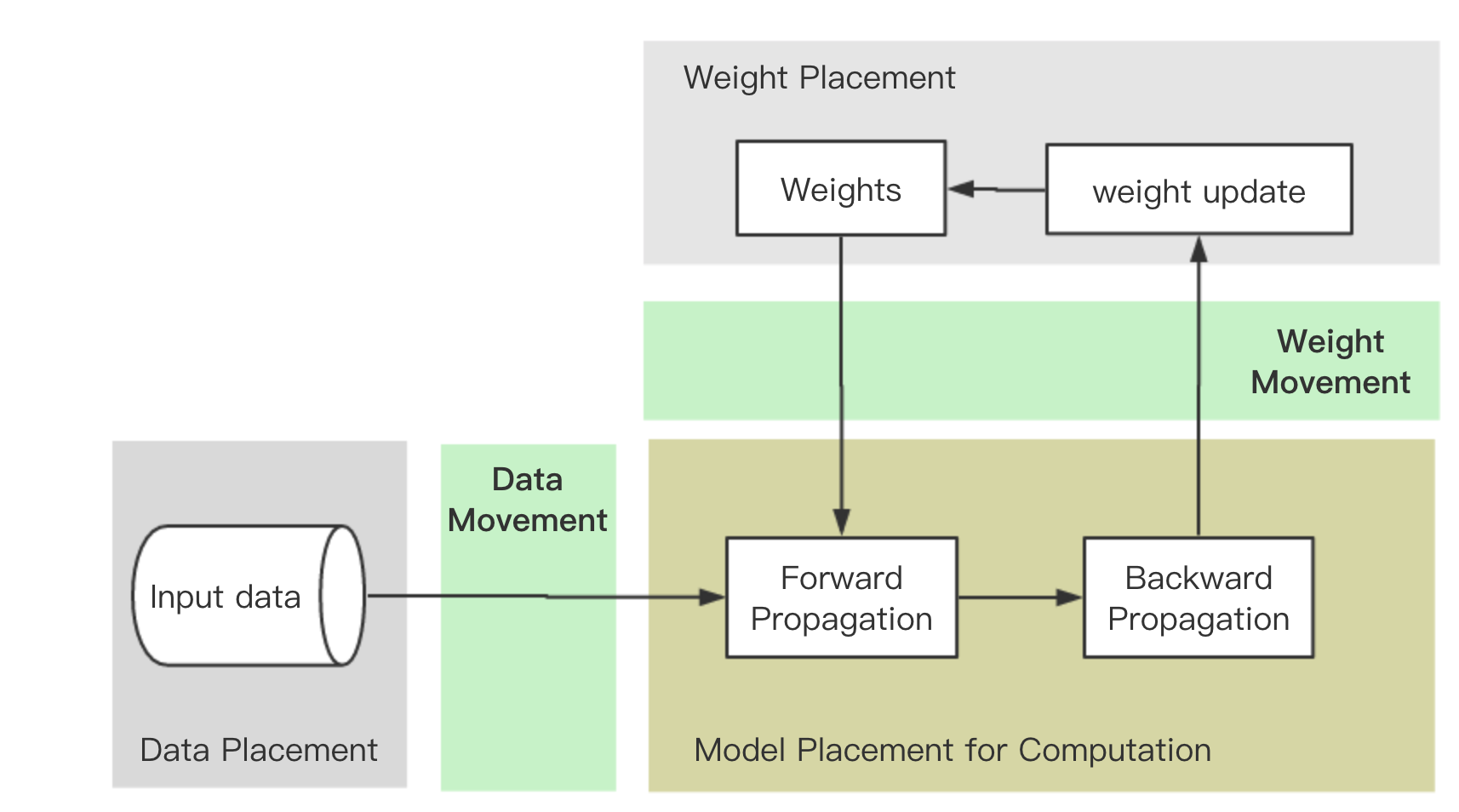}
	\vspace{-0.2cm}
 	\caption{Data Flow of A Typical DL Training Step.}
 	\label{DL_computation}
 	\vspace{-0.3cm}
\end{figure}

A DL training job always runs in an iterative fashion. Fig. \ref{DL_computation} shows the basic workflow in a typical training step. 
We study impact of placement of the input data, model computation and weight update on runtime behavior of a training job.
Weight movement
refers to data transfer related to trainable parameters, including variable reading/gradient aggregation,
respectively, in forward/backward stages.
The data movement involves storage I/O, \emph{i.e.}, feeding training samples.
For GPU workloads, the main computation is placed on GPUs while the data is in the CPU memory; therefore, input data I/O involves traffic on the CPU-GPU interconnect, \emph{i.e.}, PCIe. 

Previous workload characterization work \cite{adolf2016fathom,li2018tartan} mainly focus on measuring the relationship between model computation and weight movement traffic, while ignoring the data part. However, we found that data I/O is a non-negligible factor 
for the runtime performance, especially for single-node training workloads.


We denote the non-distributed training workloads as \emph{1w1g} (single-worker-single-GPU), and classify our distributed training workloads into four types: 
\begin{itemize}
	\item \emph{1wng}: centralized training placed locally within a single server. Typically the parameters are placed on CPU while the computation model is replicated across multiple GPUs.
	\item \emph{PS/Worker}: PS training framework 
	with each worker/PS node being placed on a separate server. 
	\item \emph{AllReduce-Local}: \emph{AllReduce} workloads in the local mode,  
	running on individual servers equipped with NVLink to exploit the high-speed multi-GPU interconnect.
	\item \emph{AllReduce-Cluster}: \emph{AllReduce} workloads running across multiple servers. 
\end{itemize}

\begin{table}[!htbp]
\vspace{-0.3cm}
\caption{Summary of five types of workloads in our cluster.}
\label{table_Summary}
\scriptsize
\centering
\begin{tabular}{ c |c| c |c  }
\hline
& System & System  & Weight    \\
& Architecture & Configuration &  Movement   \\
\hline  \hline
1w1g &-  & Local & - \\
\hline
1wng & Centralized & Local & PCIe \\
\hline
PS/Worker  & Centralized & Cluster & Ethernet \& PCIe \\
\hline 
AllReduce-Local & Decentralized & Local & NVLink \\
\hline
AllReduce-Cluster & Decentralized & Cluster & Ethernet \& NVLink \\
\hline
\end{tabular}
\vspace{-0.2cm}
\end{table}

Table \ref{table_Summary} summarizes the basic features for each type of workloads. The common features among different types of workloads are not listed.
For example, for all types model computation is placed on GPUs and the input data I/O is via PCIe from CPU to GPUs.


\subsection{Workload Characterization Framework}
\label{sec_workflow}
\begin{figure*}[!htb]
\vspace{-0.4cm}
\centering
	\centerline{\includegraphics[width=\linewidth]{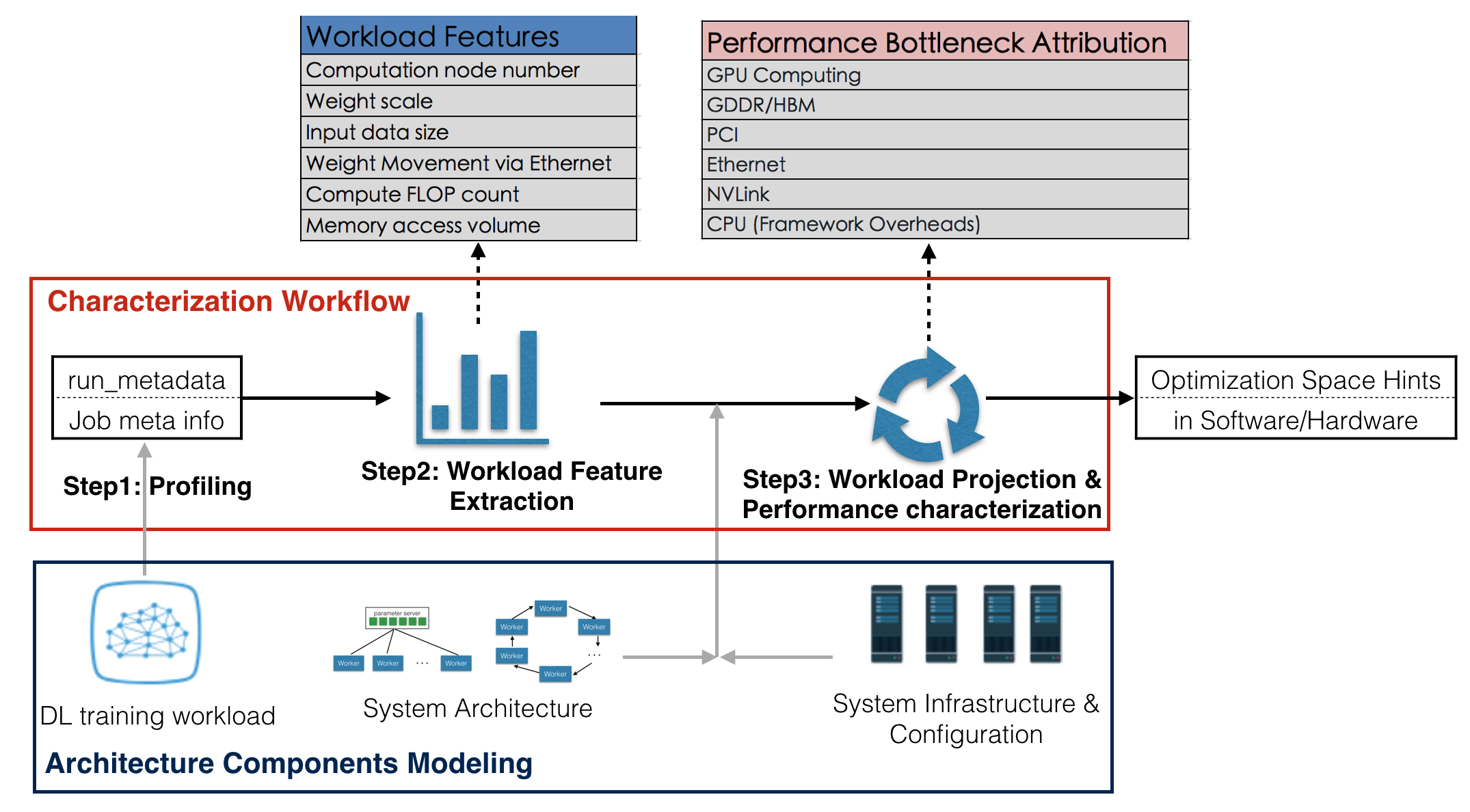}}
	\vspace{-0.4cm}
 	\caption{Workload Characterization Framework.}
 	\label{analysis_overview}
 	\vspace{-0.4cm}
\end{figure*} 
To analyze workload performance on our cluster, we established a workload characterization framework, as shown in Fig. \ref{analysis_overview}. 

\subsubsection{Runtime Profiling}
TensorFlow provides a basic profiling tool, \emph{tf.RunMetadata()} \cite{goldsborough2016tour}, which can trace the runtime information including device placement, operation attributes,
kernel launch \& execution time, and tensor attributes (data type, shape, allocation time and liveness, \emph{etc}). 

We further collect the \emph{job meta information}, which mainly includes the resource allocation information in the entire job. 
For example, for a distributed training job in the \emph{PS/Worker} architecture,
\emph{run\_metadata} provides behavior of a single computation node (using one GPU device),
 and the job meta information provides supplementary information such as how many workers the job uses.
Data collected through \emph{run\_metadata} and the \emph{job meta information} constitute the raw data for our workload analysis.

\subsubsection{Workload Feature Extraction}
We extract workload features from the fine-grained information collected, which characterize the execution requirements of each job in computation, I/O and weight/gradient transfer.
Our 
 workload feature schema is shown in Fig. \ref{analysis_overview}. 

\subsubsection{
Performance Breakdown}
For a given training job, we are interested in the composition of its execution time: 
 input data I/O time ($T_d$), computation time ($T_c$) and 
weight/gradient communication time ($T_w$).
In practice, sophisticated optimizations are 
possible to overlap computation and data transfer \cite{zhang2017poseidon,hashemi2018tictac}. Our goal is not to 
precisely model the total execution time, but to characterize the relative time consumption among computation, input I/O and weight/gradient communication. Therefore, 
potential overlap is not considered in our analysis and summation of all parts is used as the prediction of the total execution time for one training iteration/step: $T_{total}=T_d+T_c+T_w$.

\noindent \textbf{Input data I/O time.} $T_d$ measures the transport efficiency to load the input data,
computed as $T_d=\frac{S_d}{B_d}$,
where $S_d$ is the input data size and $B_d$ is the bandwidth for input data transfer. 

\noindent \textbf{Weight movement time.} $T_w$ 
can be estimated using $T_w=\frac{S_w}{B_w}$,
where $S_w$ denotes the weight size to be transferred across different model replicas within a training step,
and $B_w$ is the bandwidth of the communication medium.

\noindent \textbf{Computation time.}
The operations in DL workloads are divided into compute-bound and memory-bound ones.
FLOP count, denoted as $\# FLOPs$, is adopted to measure the computation requirements by compute-bound operations (e.g., convolution and MatMul).
The memory-bound operations, known as element-wise operations, spend more time on memory access, and thus the amount of memory access is used as their resource requirement. Let $S_{mem\_access}$ represent the total data size of memory access.
The computation time can be computed as the sum of the two parts: 
	\begin{equation}
	T_c=\frac{\#FLOPs}{peak\_FLOPs}+\frac{S_{mem\_access}}{B_{mem\_access}},
	\end{equation}
where $peak\_FLOPs$ and $B_{mem\_access}$
denote computation capacity and memory access bandwidth of the GPU, respectively.
In practice, $peak\_FLOPs$ and $B_{mem\_access}$/$B_d$/$B_w$ are usually not fully used by a workload. Therefore, we use 70\% of the actual capacities in the denominators when computing $T_c$/$T_d$/$T_w$ in our analysis.
How to measure the utilization more precisely will be part of our future work.


The time percentage of each component is further computed by dividing the time of each component by the total time, e.g., percentage of the input data I/O time is $\frac{T_d}{T_{total}}$. 

\section{Performance Characterization: Collective Behaviors}
\label{sec_clusterbehavior}
In this section, we conduct statistical analysis of tens of thousands of jobs running on PAI 
within the period of Dec.~1st, 2018 to Jan.~20th, 2019. The workloads are run on our internal TensoFlow framework, which is compatible with community TensorFlow 1.8. 
Due to the small amount of \emph{AllReduce} jobs within this period, we focus on the analysis of \emph{1w1g}, \emph{1wng} and \emph{PS/Worker} workloads from our cluster, and will further explore how much potential improvement can be achieved if using the \emph{AllReduce}-based decentralized architecture.


\subsection{Overview of the Workloads}
\label{subsec_overview}

\begin{figure}[!htb]
\vspace{-0.3cm}
\centering
	\begin{minipage}[b]{0.4\linewidth}
	\centerline{\includegraphics[width=\linewidth]{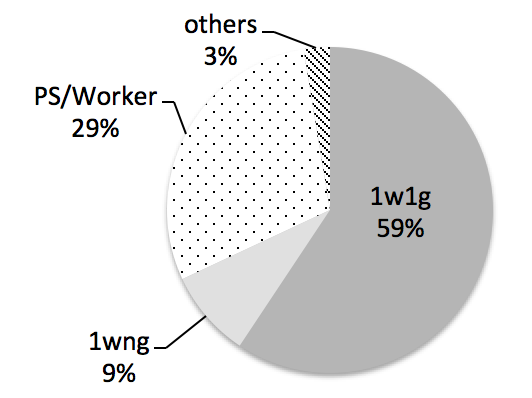}}
	\centerline{\footnotesize (a) job-level}
	\end{minipage}
	\hspace{0.3cm}
	\begin{minipage}[b]{0.4\linewidth}
	\centerline{\includegraphics[width=\linewidth]{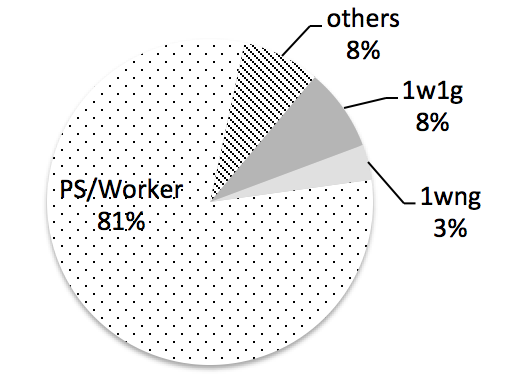}}
	\centerline{\footnotesize (b) cNode-level}
	\end{minipage}
	\caption{Constitution of Workloads.}
	\label{statistics_dist_strategy}
	\vspace{-0.3cm}
\end{figure}
Composition of different types of workloads is shown in Fig. \ref{statistics_dist_strategy}. Besides job numbers, we also count the numbers of computation nodes. A computation node, or cNode, is 
a GPU device holding a single computation model replica. 
At job-level, \emph{1w1g} dominates the job types; after taking the cNodes number in jobs into consideration, \emph{PS/Worker} jobs consume the largest portion of resources, up to 81\%.

\begin{figure}[!htb]
\centering
	\begin{minipage}[b]{0.48\linewidth}
	\centerline{\includegraphics[width=\linewidth]{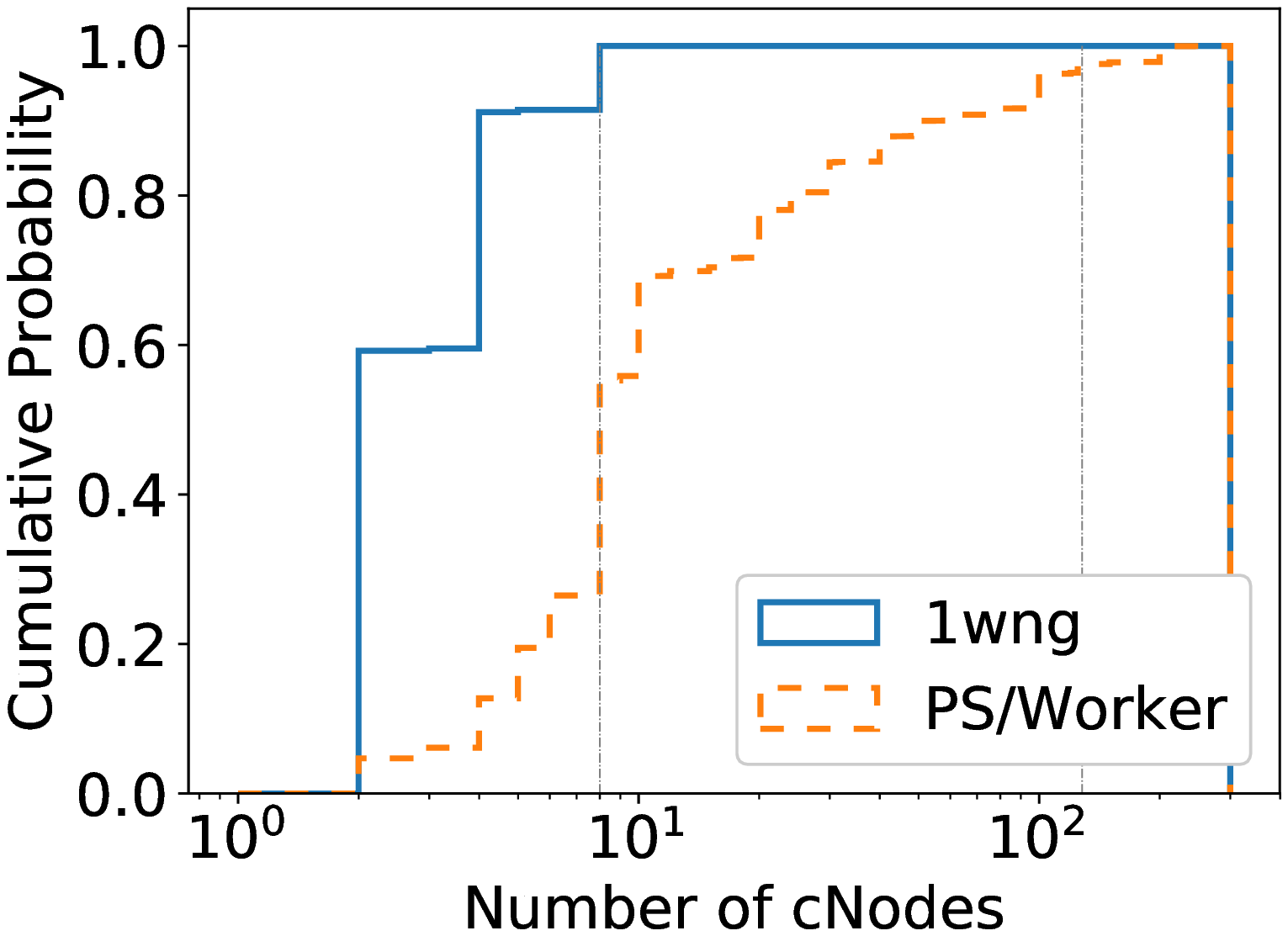}}
	\centerline{\footnotesize (a) computation scale}
	\end{minipage}
	\begin{minipage}[b]{0.48\linewidth}
	\centerline{\includegraphics[width=\linewidth]{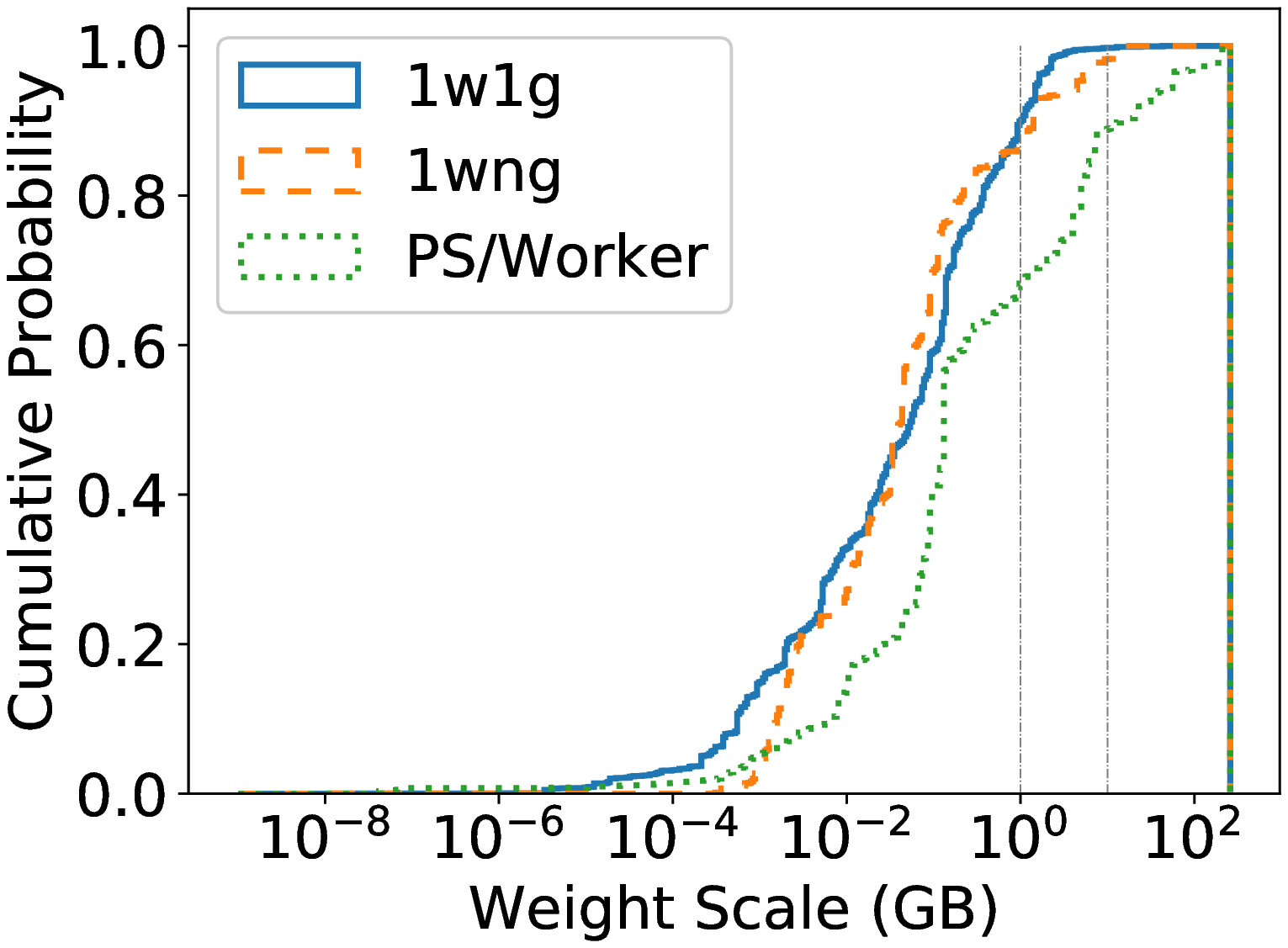}}
	\centerline{\footnotesize (b) weight size}
	\end{minipage}
	\vspace{-0.2cm}
 	\caption{Workload Scale Distribution.}
 	\label{fig_workload_scale}
 	\vspace{-0.5cm}
\end{figure} 

We further show the cumulative distribution function (CDF) of the cNode number in each type of workloads in Fig. \ref{fig_workload_scale}(a).
For \emph{1w1g} workloads, the number of cNode is always 1; 
 \emph{1wng} workloads are placed within a physical server
typically, the number of cNodes is no more than 8; about half of \emph{PS/Worker} workloads are placed on more than 8 cNodes,
while a small fraction of jobs on more than 128 cNodes.
This can help explain why there is only 29\% workloads using the \emph{PS/Worker} architecture, but the percentage of cNodes they consume is up to 81\%.

The amount of computation resources consumed by a job can reflect the problem scale and may also indicate the commercial value of the workload.
In our cluster, commodity embedding, search and recommendation workloads have large training datasets and may exploit hundreds to thousands of workers to achieve high throughput on the huge training dataset.
Notably, such extra large-scale workloads always have significant commercial impact on the company's business;
however, they are often not included in DL workload benchmarks \cite{adolf2016fathom,zhu2018benchmarking}.
We find that they are non-negligible: 
only 0.7\% of all workloads have more than 128 cNodes; 
however, they consume more than 16\% computation resource on our cluster. 
In the following Sec. \ref{sec_casestudy},
we will explore the characteristics of such large-scale workloads using two example jobs in detail.

The model size in a job is a key factor to decide what system architecture is best for the job. 
For example, for small to medium scale models that can fit into the GPU memory entirely, the \emph{AllReduce-Local} configuration can be adopted, with better performance while using less system resources.
When the weight size is large (ranging from tens to hundreds of GB),
\emph{PS/Worker} architecture should be adopted to partition the variables among multiple \emph{PS} nodes (note that only weight-replica mode is supported in \emph{AllReduce} implementation in representation DL frameworks).
Fig. \ref{fig_workload_scale}(b) illustrates the weight size distribution. We can observe that, within \emph{PS/Worker} workloads, some jobs have large weight size, more than 10 GB or even 100 GB; however lots of them have quite small model sizes. So why do they choose to adopt the \emph{PS/Worker} architecture? Can they be further optimized using a better model placement and system architecture?  We will answer these questions in Sec. \ref{subsec_opt_space}.

\begin{figure}[!htb]
\vspace{-0.3cm}
\centering
	\centerline{\includegraphics[width=\linewidth]{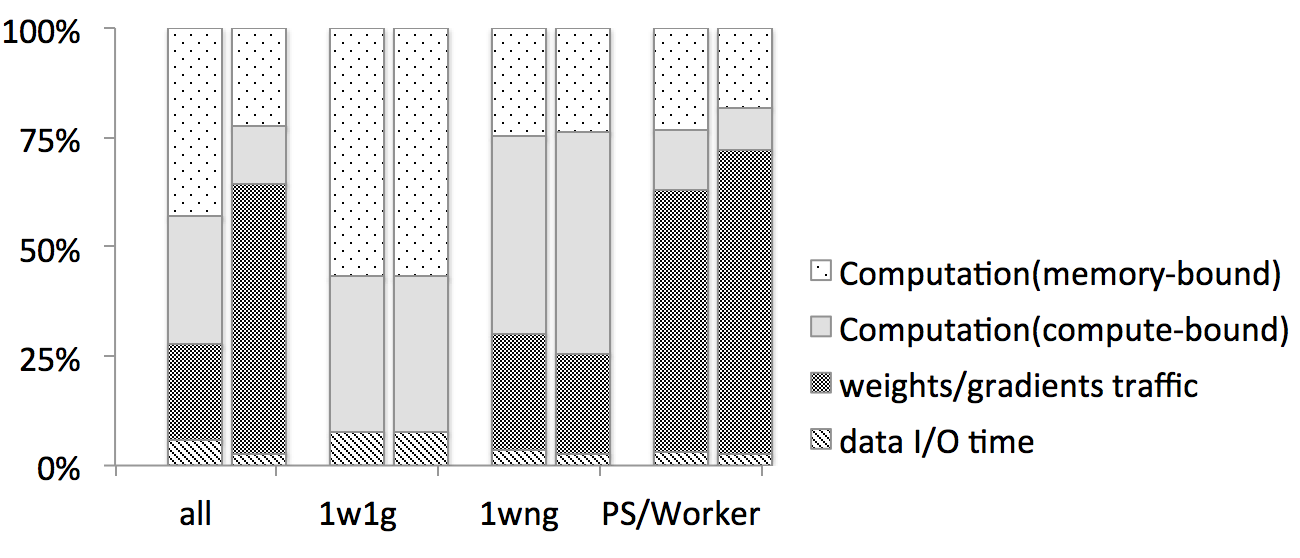}}
	\vspace{-0.3cm}
 	\caption{Average percentage of different parts of workload execution time. \emph{Left column: job-level, Right column: cNode-level.}}
 	\label{percent_type_jobs}
 	\vspace{-0.3cm}
\end{figure}


\begin{figure*}[!htb]
\vspace{-0.3cm}
	\begin{minipage}[b]{0.24\linewidth}
	\centerline{\includegraphics[width=\linewidth]{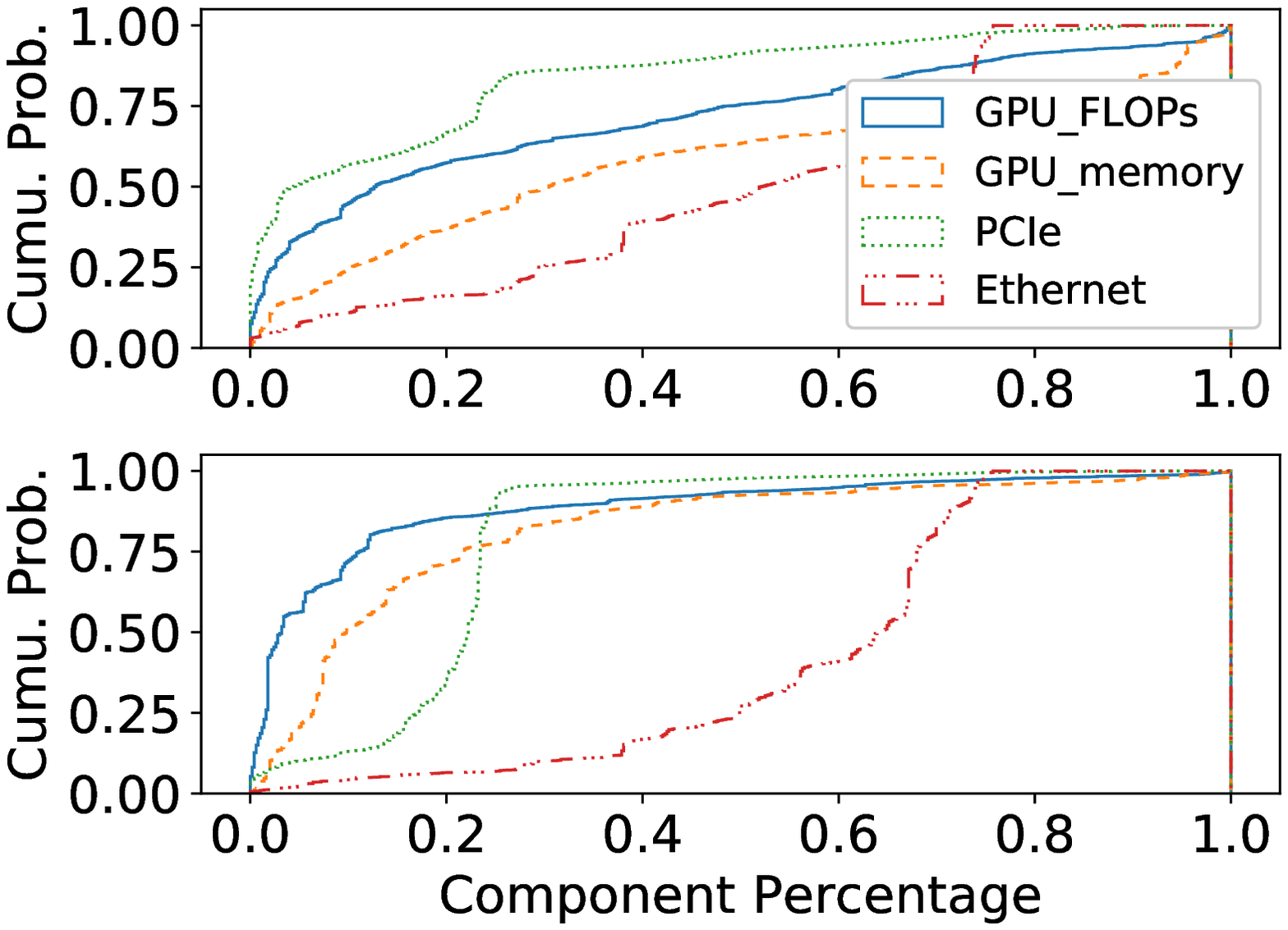}}
	\centerline{\footnotesize (a) all}
	\end{minipage}
	\begin{minipage}[b]{0.24\linewidth}
	\centerline{\includegraphics[width=\linewidth]{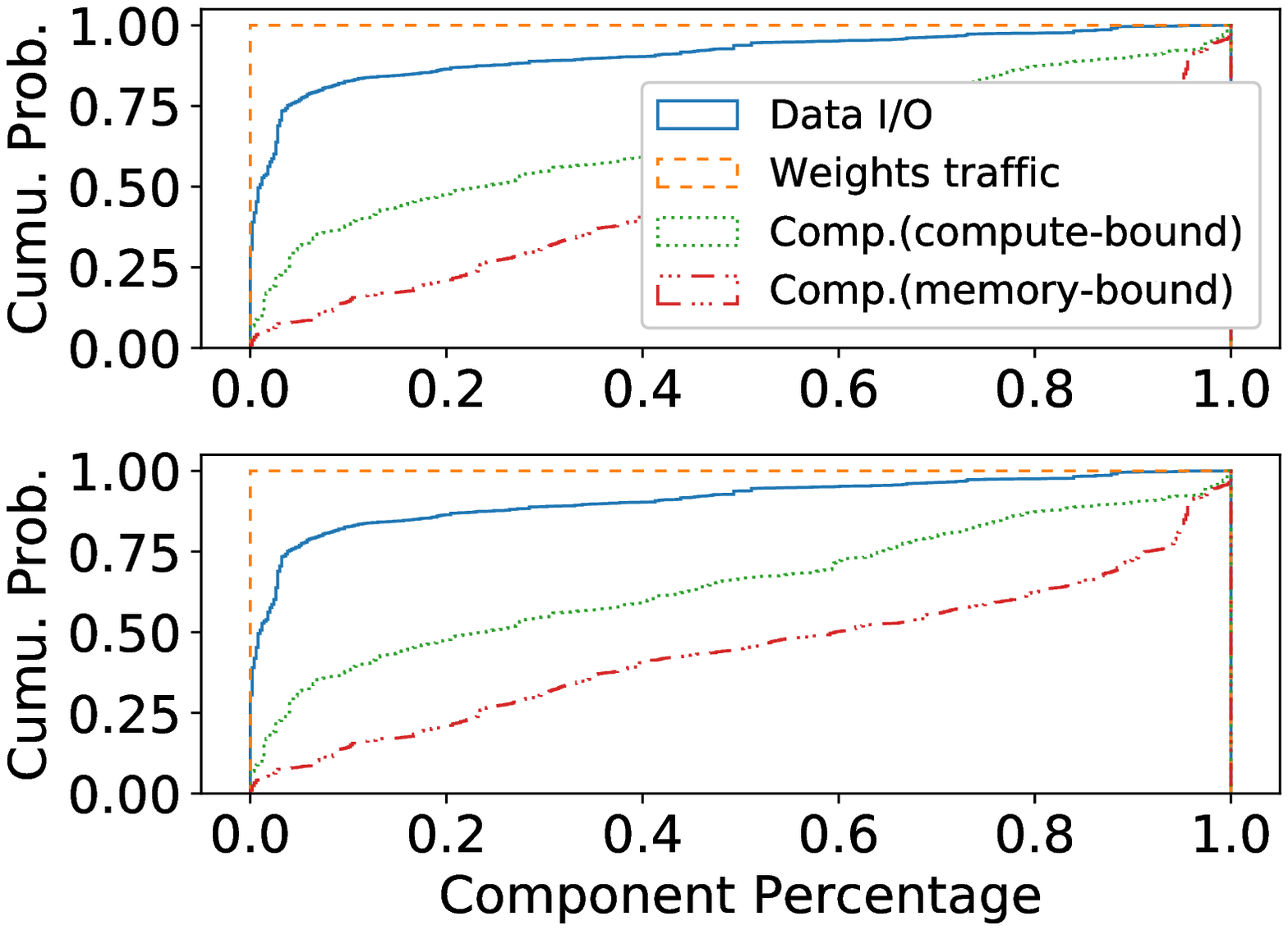}}
	\centerline{\footnotesize (b) \emph{1w1g}}
	\end{minipage}
	\begin{minipage}[b]{0.24\linewidth}
	\centerline{\includegraphics[width=\linewidth]{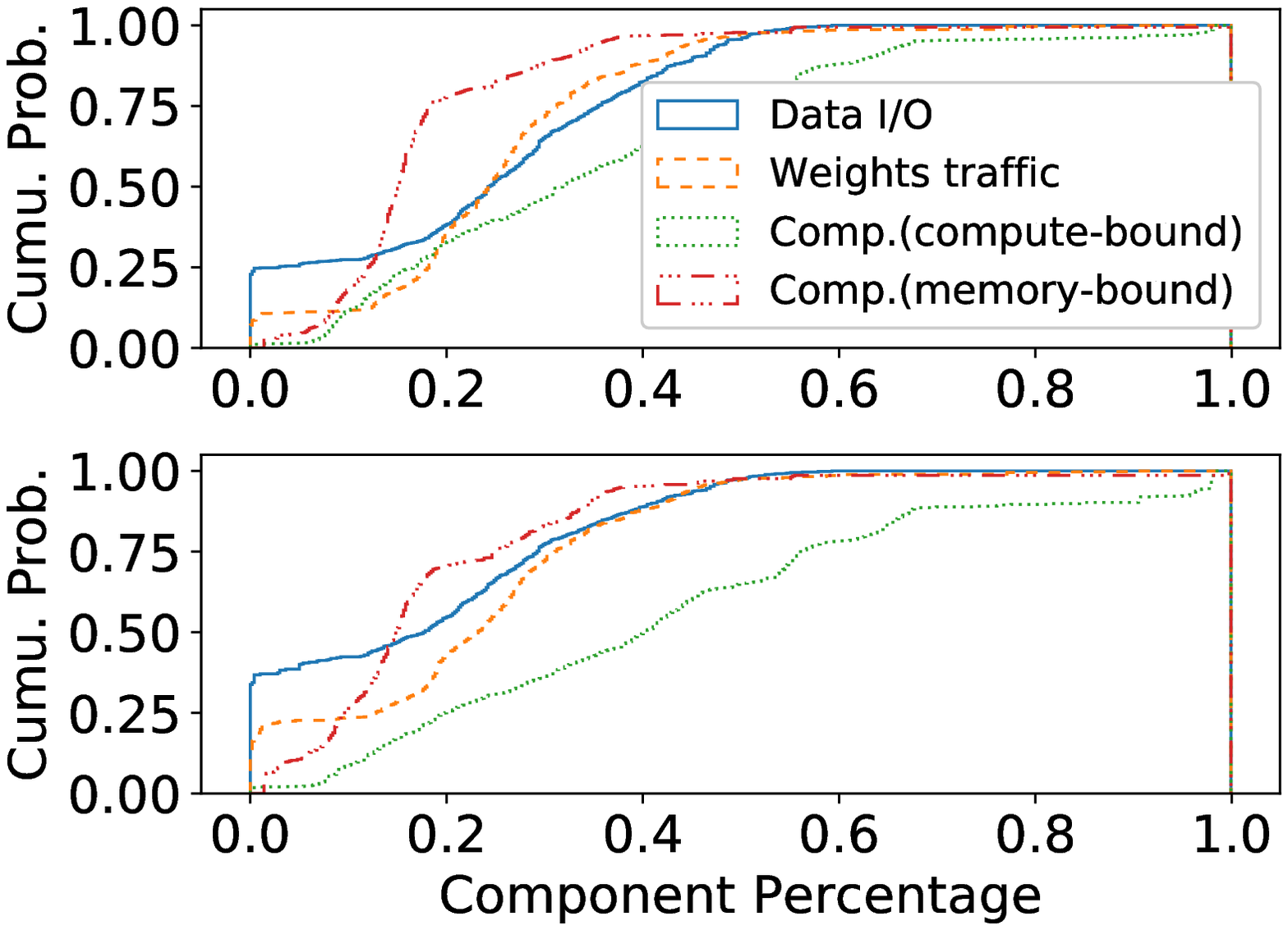}}
	\centerline{\footnotesize (c) \emph{1wng}}
	\end{minipage}
	\begin{minipage}[b]{0.24\linewidth}
	\centerline{\includegraphics[width=\linewidth]{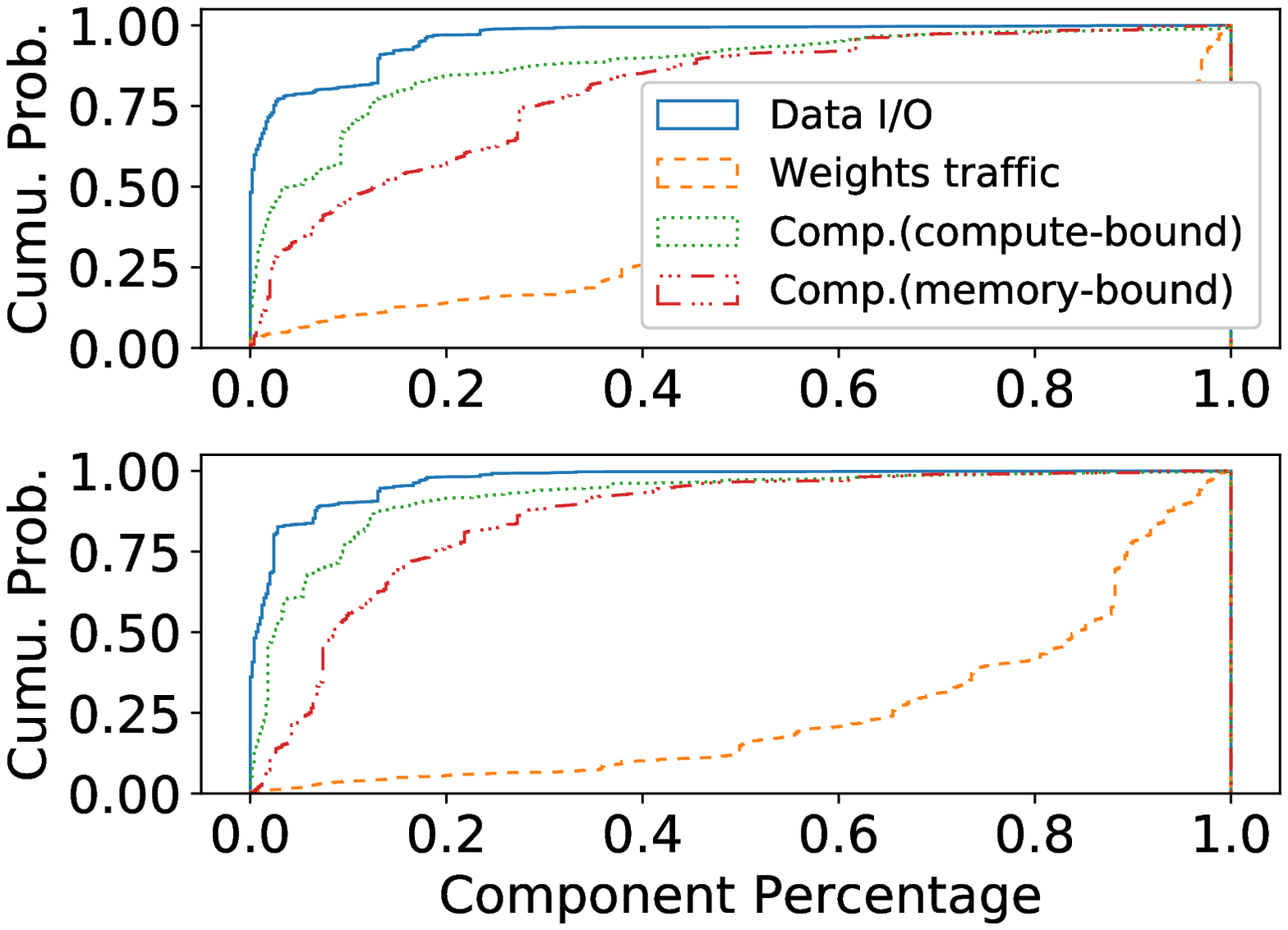}}
	\centerline{\footnotesize (d) \emph{PS/Worker}}
	\end{minipage}
	\vspace{-0.2cm}
	\caption{CDF of each component of the execution time among different workloads. \emph{Top: CDF at job-level, down: cNode-level.}}
	\label{fig_CDF_type_jobs}
	\vspace{-0.3cm}
\end{figure*}

\subsection{Performance Breakdown}
\label{subsec_runtime_breakdown}
Figure \ref{percent_type_jobs} shows the execution time breakdown  for various 
workloads,  
including time for input data I/O, weight/gradient transfer and computation.
The cNode-level percentages are computed as weighted sum of the job-level percentages, with the weight being the cNode number of each job over the overall cNode number.
Please note that \emph{1w1g} jobs do not need weight/gradient communication.
Figure \ref{fig_CDF_type_jobs} shows the detailed CDF of each component of the execution time, among the jobs. 
\mengdi{As the mapping from execution time components to  hardware differs in different types of workloads (such as the weight movement is carried out via different hardware as shown in Table \ref{table_Summary}),
we summarize the overall time breakdown according to time spent on different hardware components and show the results in Fig. \ref{fig_CDF_type_jobs}(a).}


\textbf{Input Data I/O.}
Figure \ref{fig_CDF_type_jobs} shows that for \emph{1wng} and \emph{PS/Worker} workloads,
input data movement time can be nearly ignored, approximately about 3\% on average, partially because the weight/gradient transfer time is too large. 
One thing to note is that when such workloads are mapped to another system architecture
or using a different hardware configuration, 
 the bottleneck may shift, exposing the data I/O part, which will be illustrated in Sec. \ref{subsec_opt_space}. 

For \emph{1w1g} workloads, the data I/O part is about 10\% on average.
Especially, there are about 5\% of the workloads spending more than 50\% time on input data movement,
in which case the data I/O load on PCIe becomes the bottleneck.

\textbf{Weight/Gradient Transfer.}
On average, weight/gradient communication contributes approximately 22\% to the total execution time. 
When evaluating the percentage in the cNode-level,
the proportion will be more than 60\%,
indicating that workloads with larger 
cNode numbers
are more likely to suffer from the communication bottleneck.
This can also be shown from the CDF of time breakdown of \emph{PS/Worker} workloads in Fig. \ref{fig_CDF_type_jobs}(d). The \emph{PS/Worker} workloads always involve large numbers of cNodes
with large proportions of time spent on weight/gradient transfer.
Specifically, more than 40\% \emph{PS/Worker} jobs 
spend more than 80\% time in communication via Ethernet and/or PCIe.
Given the high communication overhead, a potential improvement to expedite model training is to upgrade the network facility or to vary the system configuration by porting the \emph{PS/Worker} workloads to \emph{AllReduce-Local} for leveraging the high communication efficiency introduced by NVLink.


\textbf{Computation.}
Computation can be further decomposed into memory-bound and compute-bound 
computation.
We can see that memory-bound computation time is larger than compute-bound operation time in all types of workloads.
This indicates that the workloads in our cluster involve more memory access. In this case, XLA may provide powerful optimization for element-wise operations (major contribution for memory access).
XLA is a domain-specific compiler for linear algebra that optimizes TensorFlow computation, which can fuse pipelined operations to reduce the memory overhead.

Additionally, for compute-bound operations,
mixed-precision computation can also be introduced to exploit the computation power provided by TensorCore
\cite{volta}, which provides up to 8X higher peak FLOPS on Tesla V100, 
as compared to using standard FP32 operations on V100.

\subsection{Exploring the Optimization Space}
\label{subsec_opt_space}
Previously we showed a holistic execution profile for
all workloads. But how would this execution profile change under different system settings?
For instance, what can we get by upgrading the network bandwidth from 25Gbps to 100Gbps? 
Is there any further end-to-end performance speed-up by boosting the GPU 
peak computing power to 64 or 256 TFLOPS? 
Will the performance bottleneck shift to data movement by increasing GPU 
memory bandwidth to 4TB per second? In addition, what if we use \emph{AllReduce-Local} or \emph{AllReduce-Cluster} to run the PS jobs? 


Next, we analytically evaluate potential performance impact by switching the PS workloads to AllReduce and by changing system
configurations for different types of workloads.
Especially, we estimate how the performance will be like when GPUs are upgraded to more powerful ones, and interconnections are varied among PCIe (for CPU-GPU/GPU-GPU communication), 
Ethernet (for cross-server communication), and NVLink (for high-speed inter-GPU 
communication within a single machine), by changing the values of $S_d$/$S_w$/$S_{mem_{access}}$/$peak_{FLOPs}$ in the analytical models in Sec. \ref{sec_workflow}, respectively.
Tallent \emph{et al.} \cite{tallent2017evaluating} compared workload performance for GPU interconnect with NVLink and Cirrascale GX8 PCIe, and their results show that DGX-1 with NVLink has superior performance except on ResNet-type of workloads.
We would like to investigate how much the high-speed NVLink interconnect can outperform PCIe/Ethernet with our  workloads.

\subsubsection{Performance Impact of AllReduce}

Figure \ref{percent_type_jobs} shows that communication consumes an important portion of the execution time in \emph{PS/Worker} workloads, which may partially be due to the limited bandwidth of Ethernet/PCIe.
We estiamte the performance when PS workloads training small to medium scale models (that can be fit into the GPU memory entirely) are ported to the \emph{AllReduce} architectures, to exploit 
 the high-speed NVLink. 
In addition to single node performance,
we further evaluate the overall throughput of a training job, 
which can be computed as
\begin{equation}
throughput=\frac{\#cNode}{T_{total}}\times batch\_size
\label{eq_throughput}
\end{equation}
Here $\frac{\#cNode}{T_{total}}$ is the number of steps the job can train in unit time with all its computation nodes.
Considering that $batch\_size$ remains the same in each computation node, the throughput is related to 1) single-node performance $T_{total}$ and 2) the number of cNodes $\#cNode$. 

We map the \emph{PS/Worker} workloads to the \emph{AllReduce-Local} architecture as follows, since an \emph{AllReduce-Local} job can have at most 8 $\#cNodes$: 
 for a \emph{PS/Worker} job with $\#cNodes> 8$, the number of cNodes is reduced to 8; for those with $\#cNodes\le8$, the cNode numbers will remain unchanged.
To map the \emph{PS/Worker} workloads to the \emph{AllReduce-Cluster} architecture, we retain the original number of cNodes in the jobs. 
In addition to the speedup of all workloads, 
we select workloads whose throughput cannot be improved by \emph{AllReduce-Local} and show the performance acceleration with \emph{AllReduce-Cluster}.

\begin{figure}[!htb]
\vspace{-0.3cm}
	\begin{minipage}[b]{0.48\linewidth}
	\centerline{\includegraphics[width=\linewidth]{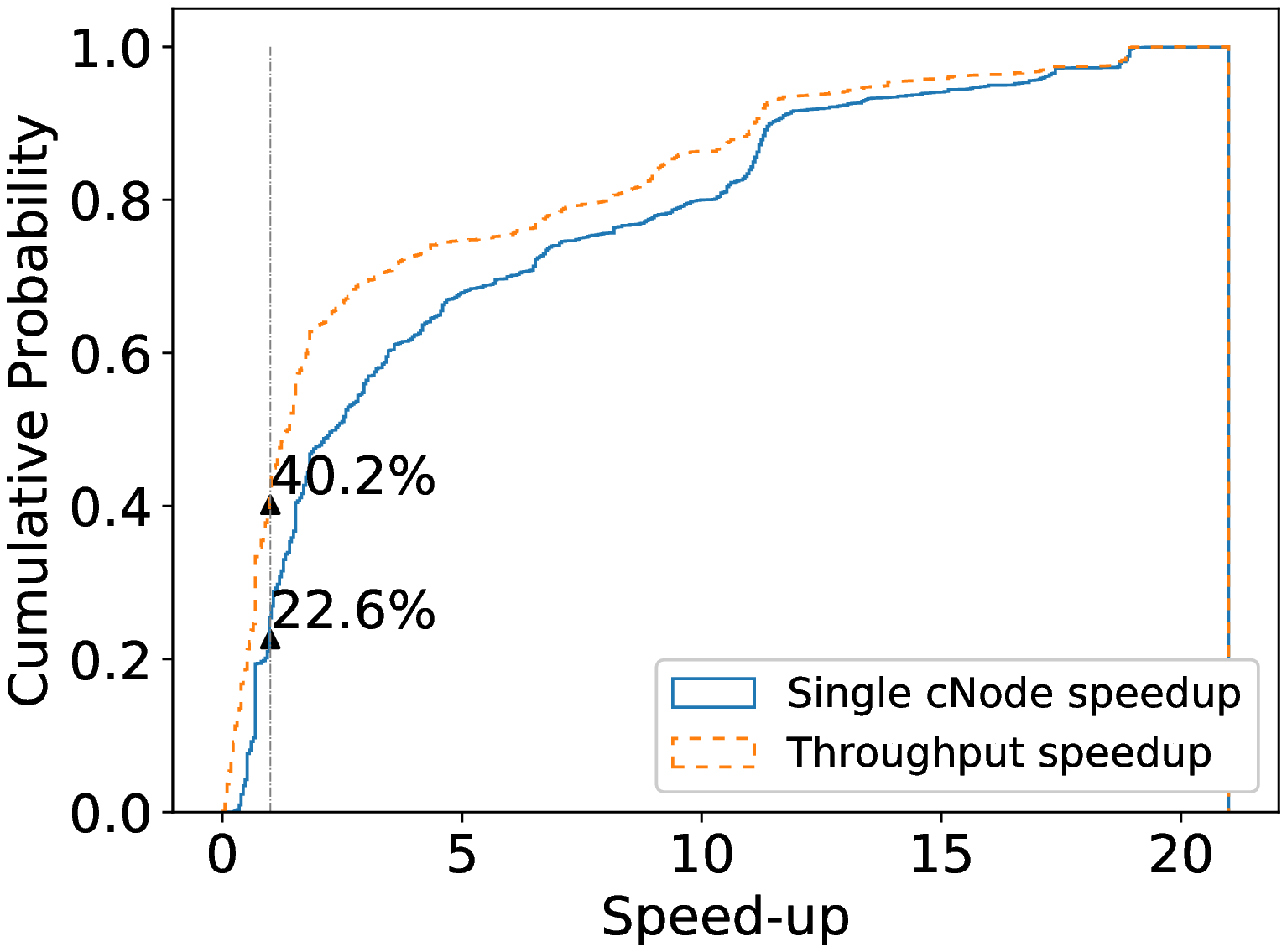}}
	\centerline{\footnotesize (a) \emph{AllReduce-Local}}
	\end{minipage}
	\begin{minipage}[b]{0.48\linewidth}
	\centerline{\includegraphics[width=\linewidth]{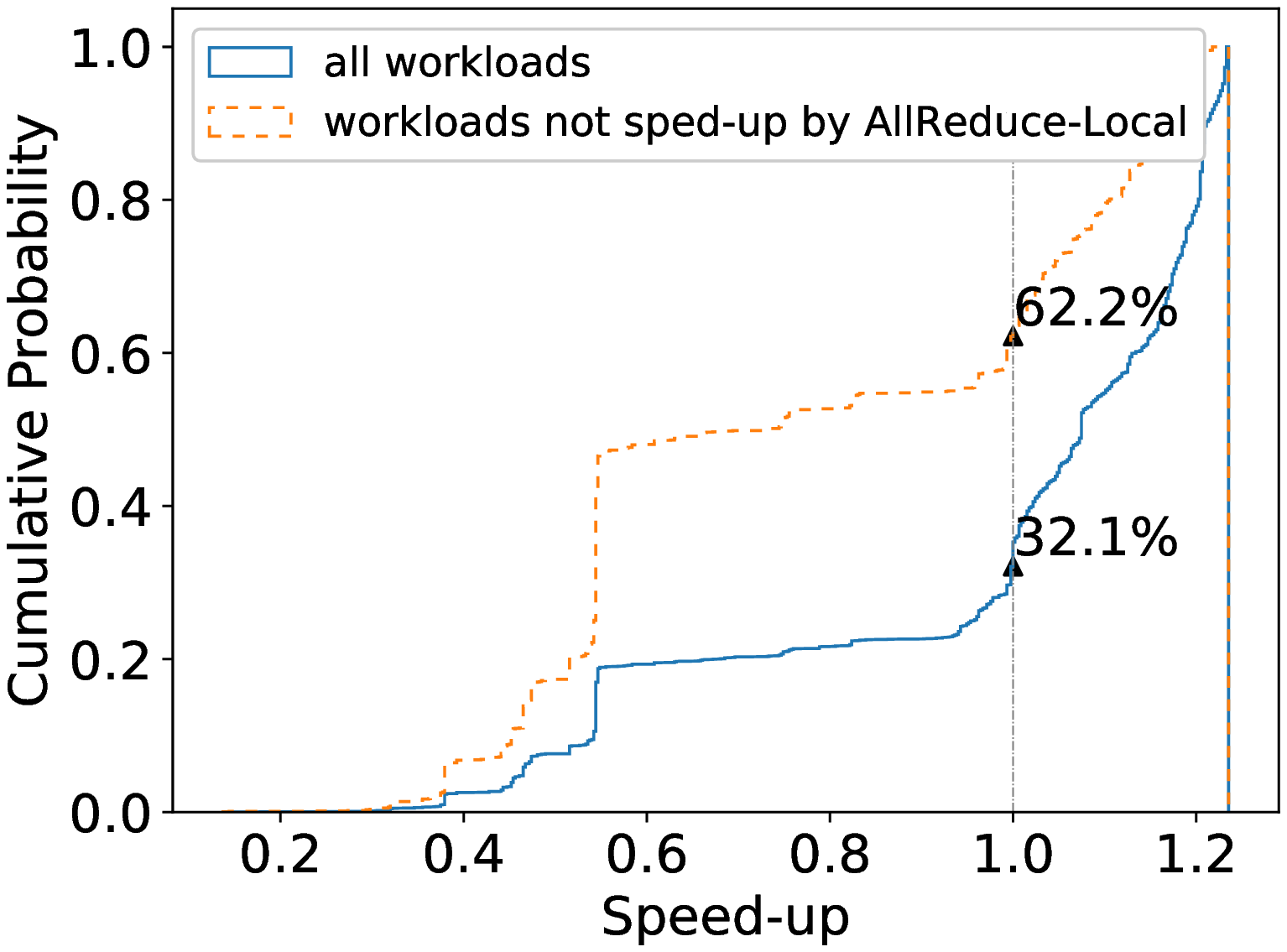}}
	\centerline{\footnotesize (b) \emph{AllReduce-Cluster}}
	\end{minipage}
	\vspace{-0.2cm}
	\caption{Improvement by mapping the workloads to \emph{AllReduce}.}
	\label{opt_GPU_NVLink}
	\vspace{-0.3cm}
\end{figure}

\begin{figure}[!htb]
\vspace{-0.3cm}
	\begin{minipage}[b]{0.61\linewidth}
	\centerline{\includegraphics[width=\linewidth]{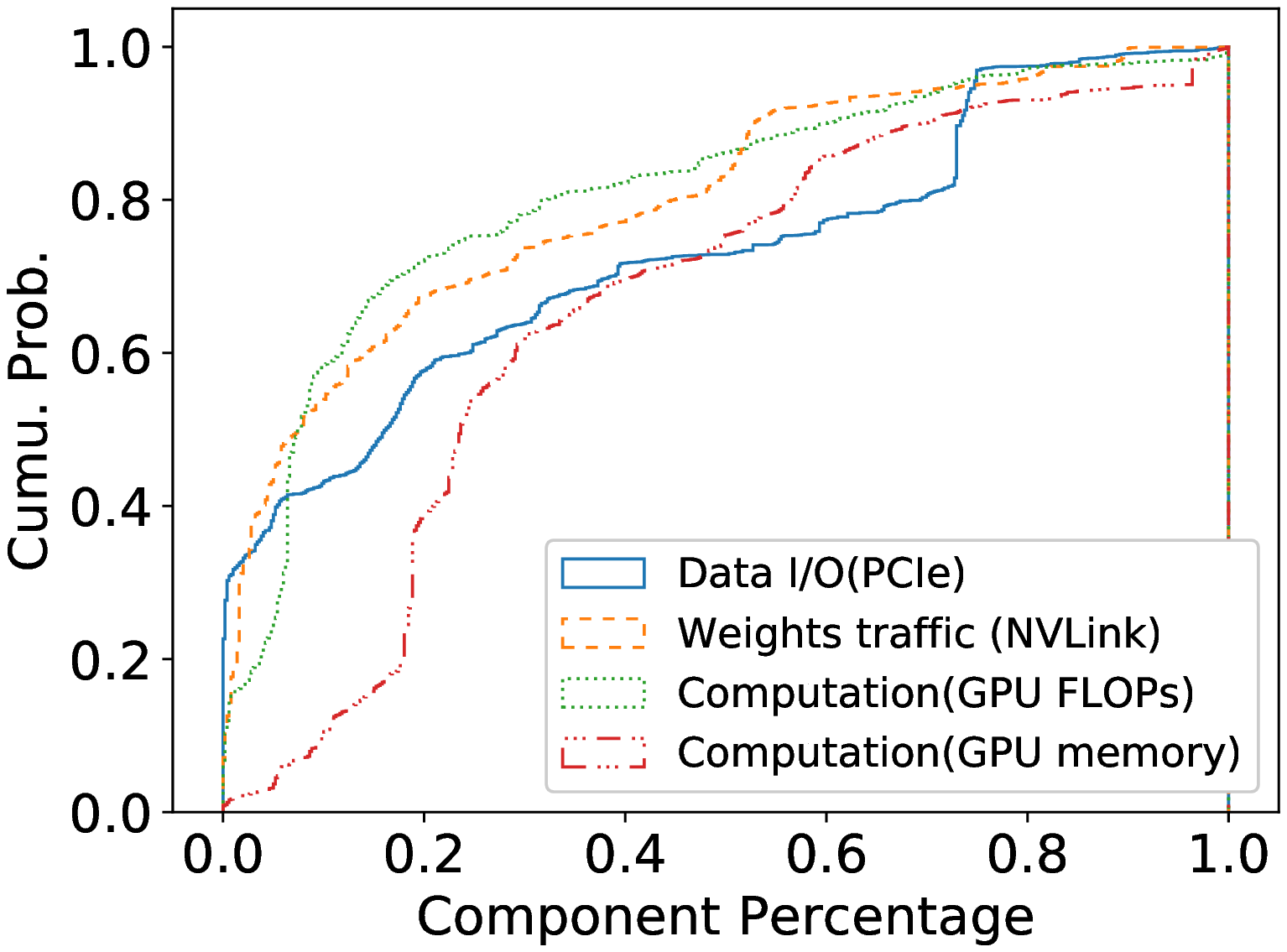}}
	\centerline{\footnotesize (a) CDF}
	\end{minipage}
	\begin{minipage}[b]{0.31\linewidth}
	\centerline{\includegraphics[width=\linewidth]{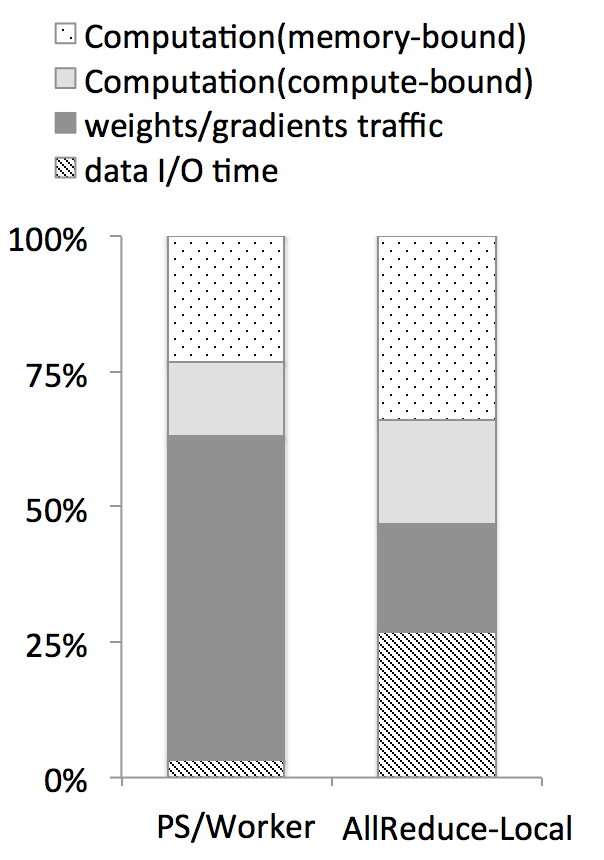}}
	\centerline{\footnotesize (b) Average Breakdown}
	\end{minipage}
	\vspace{-0.2cm}
	\caption{Performance breakdown of \emph{PS/Worker} workloads after being mapped to \emph{AllReduce-Local}.}
	\label{fig_breakdown_allreduce}
	\vspace{-0.4cm}
\end{figure}

Fig. \ref{opt_GPU_NVLink} shows that by shifting the communication medium from PCIe/Ethernet to the high speed NVLink interconnect with \emph{AllReduce-Local},
most of the workloads can be accelerated 
at different levels.
Considering the potential reduction of $\#cNode$ in projection, 
about 60\% workloads still achieve speedup in the overall throughput. 
This indicates that \emph{AllReduce-Local} architecture equipped with NVLink can potentially boost performance for most of the \emph{PS/Worker} workloads, while at the same time saving system resources significantly (as the number of cNodes after projection will be no more than 8, which can be much larger before projection as shown in Fig. \ref{fig_workload_scale}(a)).
We also note that about 22.6\% \emph{PS/Worker} workloads cannot benefit from switching to the \emph{AllReduce-Local} architecture. With the switching, all workloads experience acceleration of the weight/gradient transfer, as well as slow-down of input data I/O, due to the competition for PCIe bandwidth (as input data are transferred from CPU to multiple GPUs within a server simultaneously);
whether a workload is sped up or slowed down relies on which part dominates.

To demonstrate the bottleneck shift effect, we further illustrate the execution time breakdown of the \emph{AllReduce-Local} 
workloads in Fig. \ref{fig_breakdown_allreduce}.
As compared to the CDF shown in Fig. \ref{fig_CDF_type_jobs}(d),
we can observe that the weight/gradient communication part is vastly reduced,
while the other parts,
including computation and the data I/O fraction, become more important. Especially,
based on Fig. \ref{fig_breakdown_allreduce}(b), we can see that the portion of data I/O via PCIe increases the most,
indicating the shift of bottlenecks with different architectures.

When workloads are shifted from \emph{PS/Worker} to \emph{AllReduce-Cluster},
the main speedup is due to the change of weight/gradient movement medium from Ethernet\&PCIe to Ethernet\&NVLink. However, in both sets of configurations,
Ethernet is the main bottleneck for data transfer, and thus the speedup is quite limited,
at most 1.2X based on Table \ref{table_baseline_config}.
On average, 67.9\% workloads can be sped up. Furthermore,
among the workloads that cannot be improved by \emph{AllReduce-Local},
about 37.8\% can be sped up with \emph{AllReduce-Cluster}.


\begin{table}[!htbp]
\vspace{-0.3cm}
\caption{Hardware Configuration Variations}
\label{table_provision_config}
\centering
\begin{tabular}{ c| c }
\hline
& Candidates \\
\hline
Ethernet/Gbps & \{10, 25, 100\} \\
\hline
PCI/GB & \{10, 50\} \\
\hline
GPU peak FLOPs/T & \{8, 16, 32, 64\} \\
\hline
GPU memory/TB & \{1, 2, 4\} \\
\hline
\end{tabular}
\vspace{-0.5cm}
\end{table}

\subsubsection{Performance Impact of Hardware Evolution}
\label{sec_resource_provision}
\begin{figure*}[!htb]
\vspace{-0.4cm}
	\begin{minipage}[b]{0.24\linewidth}
	\centerline{\includegraphics[width=\linewidth]{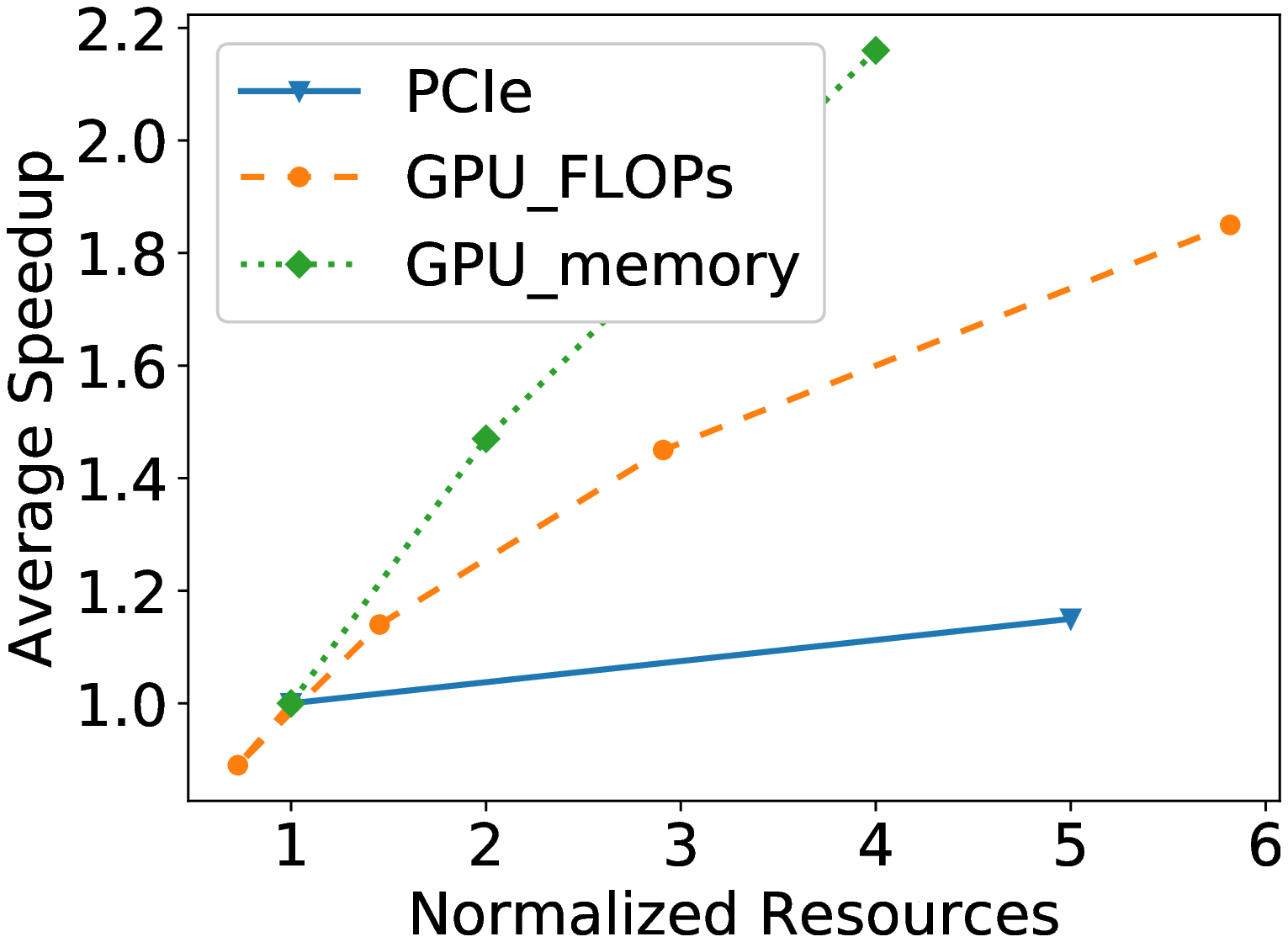}}
	\vspace{-0.2cm}
	\centerline{\footnotesize (a) \emph{1w1g}}
	\end{minipage}
	\begin{minipage}[b]{0.24\linewidth}
	\centerline{\includegraphics[width=\linewidth]{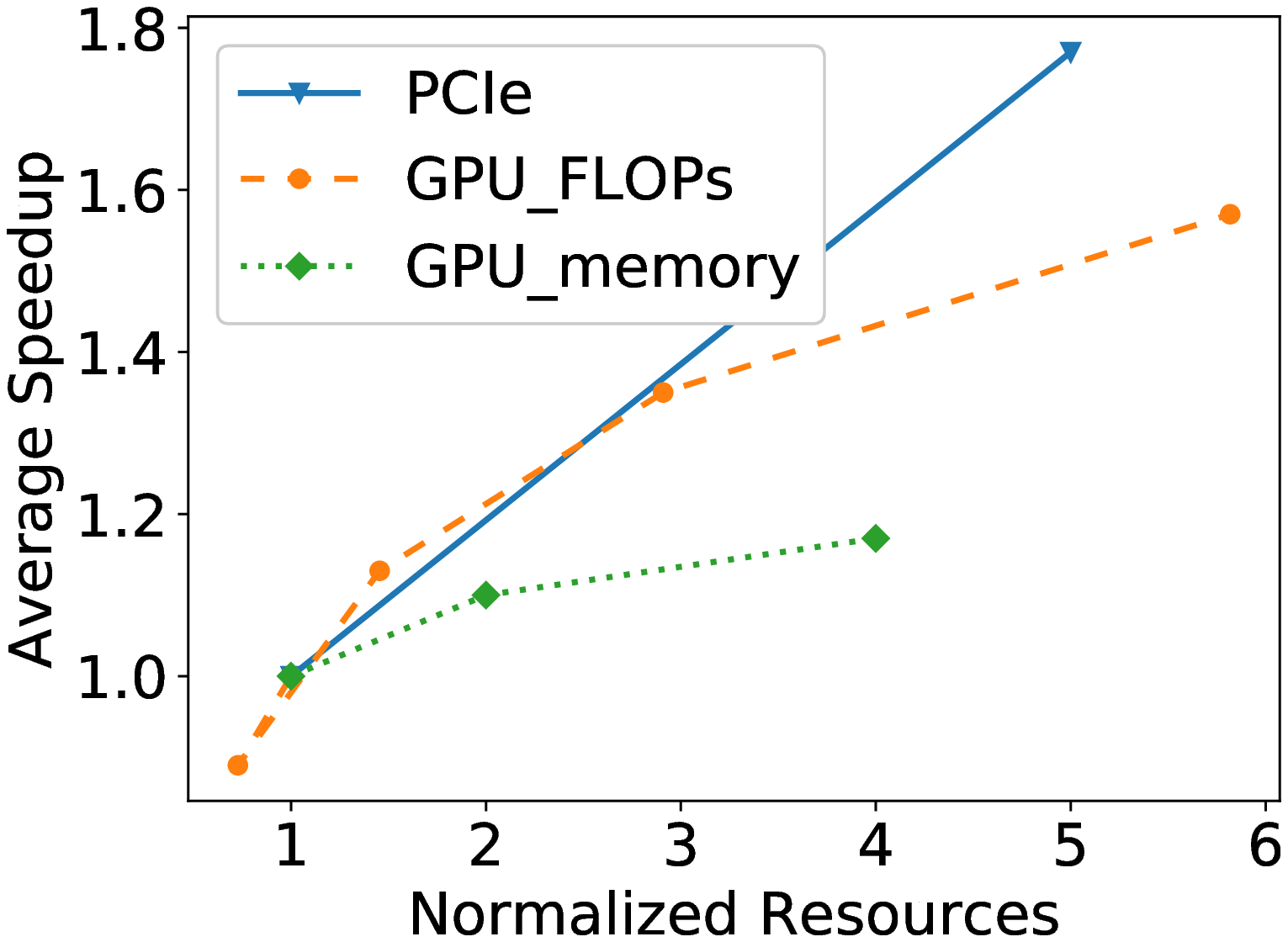}}
	\vspace{-0.2cm}
	\centerline{\footnotesize (b) \emph{1wng}}
	\end{minipage}
	\begin{minipage}[b]{0.24\linewidth}
	\centerline{\includegraphics[width=\linewidth]{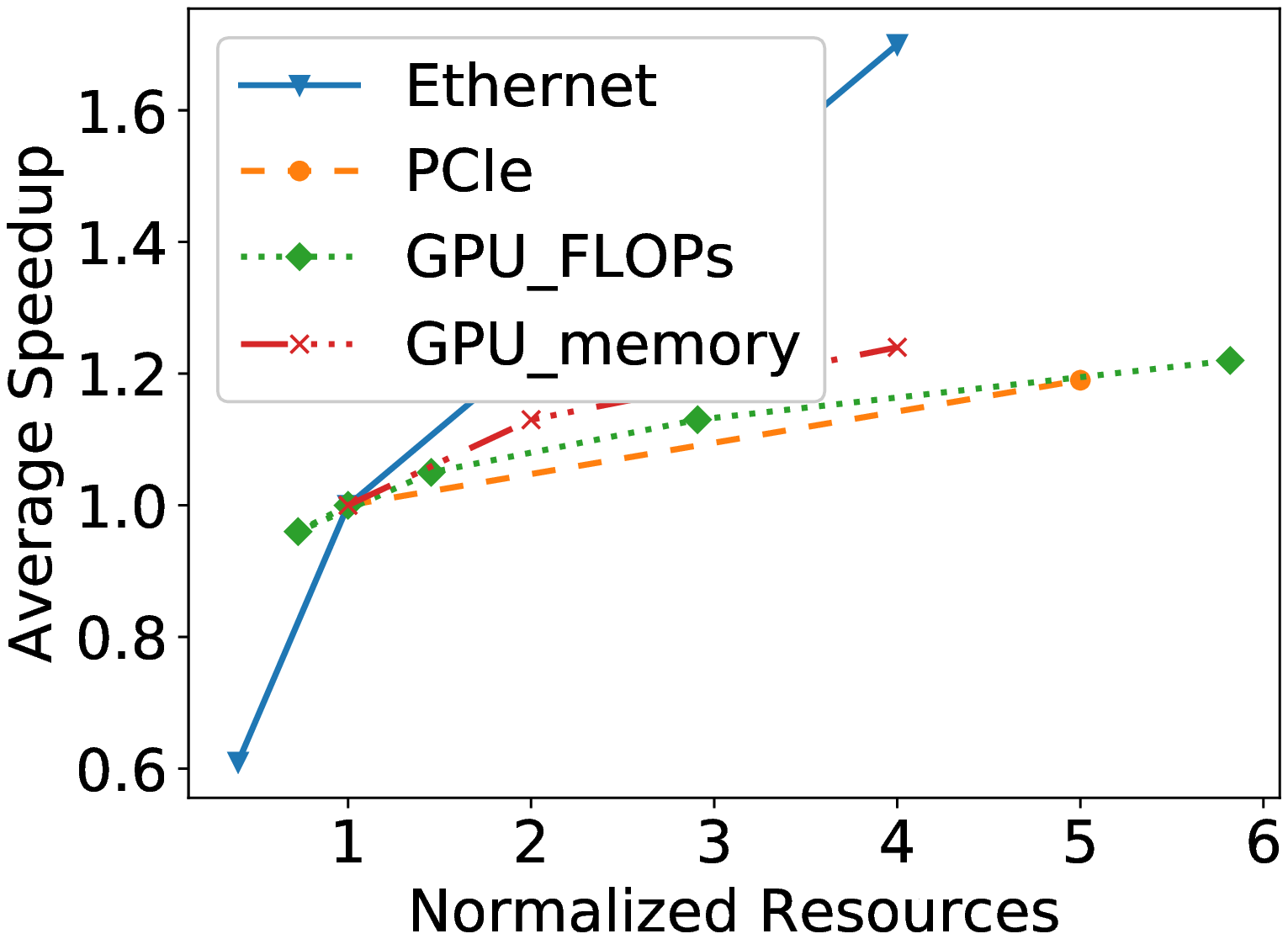}}
	\vspace{-0.2cm}
	\centerline{\footnotesize (c) \emph{PS/Worker}}
	\end{minipage}
	\begin{minipage}[b]{0.24\linewidth}
	\centerline{\includegraphics[width=\linewidth]{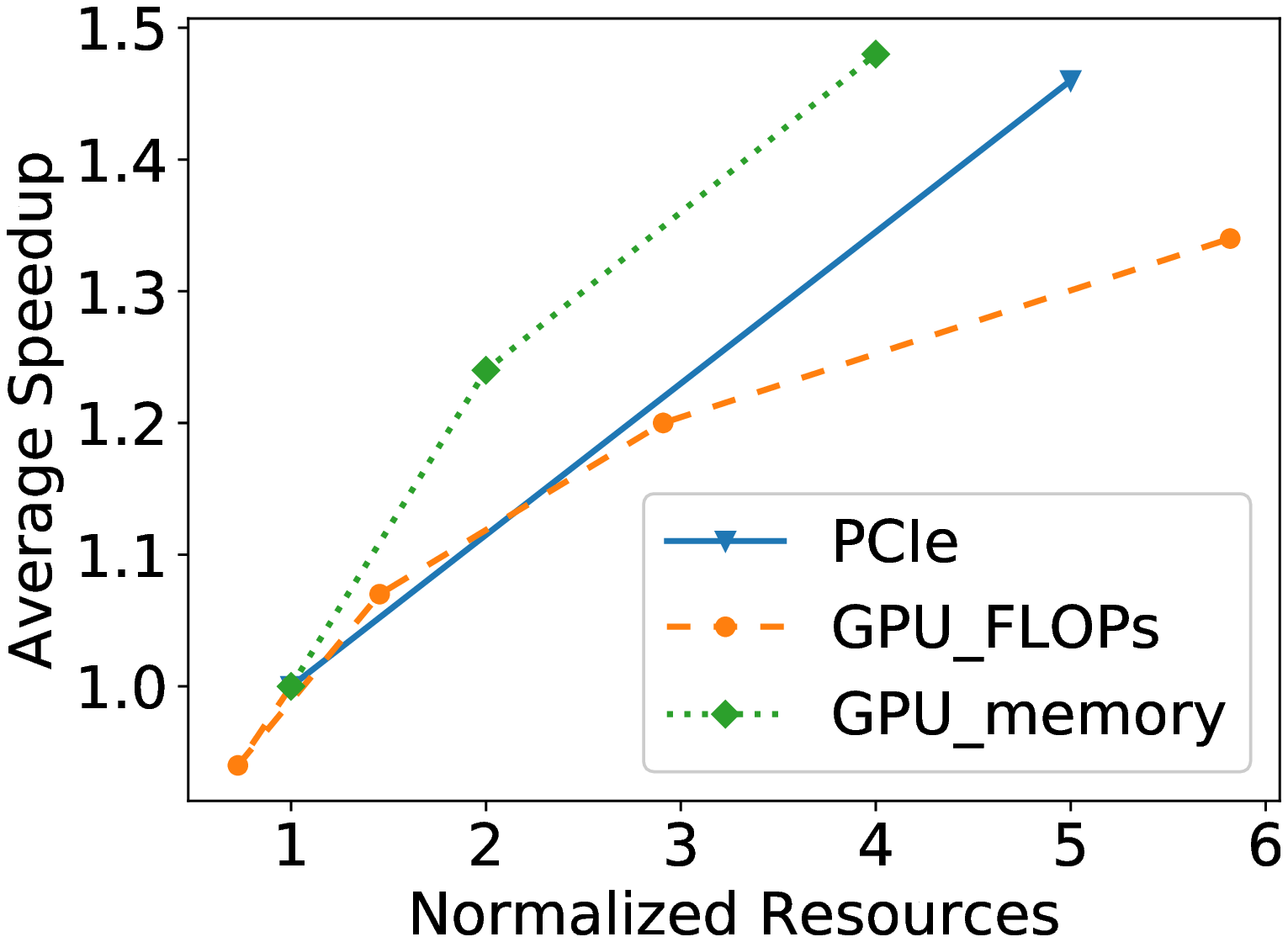}}
	\vspace{-0.2cm}
	\centerline{\footnotesize (d) \emph{AllReduce-Local}}
	\end{minipage}
	\vspace{-0.2cm}
	\caption{Speedup with different hardware configurations.}
	\label{fig_resource_provision}
	\vspace{-0.4cm}
\end{figure*}


We next investigate how the workloads perform with different hardware configurations, as shown in Table \ref{table_provision_config}. We show normalized resource values in Fig. \ref{fig_resource_provision} according to the basic settings in Table \ref{table_baseline_config}, to facilitate result comparison. For example, the Ethernet bandwidth is normalized using 25Gbps as the basic unit, and PCIe bandwidth is normalized by 10GB/s.
We evaluate both the original workloads with \emph{1w1g}, \emph{1wng} and \emph{PS/Worker} architectures, and also the mapped \emph{AllReduce-Local} workloads from the \emph{PS/Worker} workloads. 

In Fig. \ref{fig_resource_provision}, the speedup is computed using the performance achieved with the new configuration of the respective resource.  
Different workloads exhibit different behaviors:
\emph{1w1g} workloads are most sensitive to GPU memory bandwidth,
\emph{1wng} ones vary most with the variation of PCIe bandwidth,
and the \emph{PS/Worker} type relies most on the Ethernet bandwidth.
The observations are consistent with 
the performance breakdown results in Fig. \ref{percent_type_jobs} and Fig. \ref{fig_CDF_type_jobs}. 
For example, \emph{PS/Worker} workloads spend the most time on weight/gradient transfer via Ethernet and they achieve the highest speedup by improvement of the Ethernet bandwidth.

Comparison between Fig. \ref{fig_resource_provision}(c) and (d) shows the bottleneck shift effect: performance of \emph{PS/Worker} workloads varies most when varying the Ethernet bandwidth, and they are accelerated quite a bit when the GPU memory bandwidth improves; when the workloads are projected to \emph{AllReduce-Local}, GPU memory bandwidth has the largest impact on performance.

\subsection{Summary of Key Observations}
\label{subsec_key_observe}
We make several interesting observations based on the above:

$\triangleright$ 
  On PAI, distributed training jobs dominate resource consumption, with \emph{PS/Worker} jobs consuming 81\% of overall computation resources. 
 
 $\triangleright$ 90\% jobs train small-scale models, \emph{i.e.}, model size less than 10GB,
  while there exist also large-scale models(100-300GB) which are trained in large-scale distributed mode
  and consumes large amounts of resources. 
  
  $\triangleright$ On average weight/gradient communication takes almost 62\% of the total execution time
  for all workloads. 
  For \emph{PS/Worker} jobs, more than 40\% workloads spend more than 80\% time in weight/gradient communication.
  As to the computation portion which is the focus of previous studies \cite{adolf2016fathom,qi2016paleo},
  on average it only contributes 35\% of the total training time,
  with compute-bound part contributing 13\% and memory-bound part 22\%.


 $\triangleright$ Throughput of 60\% \emph{PS/Worker} workloads can be improved when they are ported to the \emph{AllReduce-Local} architecture, which can leverage the high-speed NVLink for GPU interconnect.

 $\triangleright$ Workloads show different levels of sensitivity for hardware evolution 
 and the performance bottleneck may shift with the change of system architecture. \emph{PS/Worker} workloads are most sensitive to Ethernet bandwidth; after projected to \emph{AllReduce-Local}, they benefit the most from the improvement of GPU memory access bandwidth.



\section{Performance Characterization: Case Studies}
\label{sec_casestudy}

In this section, we zoom into the training of several production DL models in detail, 
to further detect their performance bottlenecks
and evaluate several optimization techniques.

We run the selected training workloads in an experimental testbed of 64 servers. Each server is equipped with one 96-core Intel Xeon Platinum 8163 CPU, eight Tesla V100 GPUs, 128GB RAM, 10GB PCIe and 50GB NVLink. The servers are connected through 25Gbps bi-directional Ethernet. We extensively investigate data preprocessing time and the framework overhead (mostly due to CPU runtime scheduling and GPU kernel launch time),
which are not considered in Sec. \ref{sec_clusterbehavior} as they are not fundamental resource demands of workloads and can be optimized to be ignorable using different technologies. With our testbed experiments, we will show the impact of the framework overhead and discuss techniques to minimize it. 

\subsection{Selected Workloads}

Table \ref{model_config} summarizes the six models used for our case studies, selected from different application domains and with different scales of parameter size. 


\textbf{ResNet50}. Residual networks have been proven to be powerful and widely applied in multiple domains \cite{he2016identity,dai2016r}. 



\textbf{NMT}. In our production system, NMT model \cite{vaswani2017attention} has been applied to 
translation 
 for e-commerce business and others.

\textbf{Speech}. Neural acoustic models \cite{kim2017dynamic} have been useful in speech recognition and widely adopted in commercial acoustic applications.
The model we evaluate is composed of CNN followed by Long Short-Term Memory (LSTM) architecture with layer normalization.

\textbf{BERT}. BERT \cite{devlin2018bert} is one of the most commonly used 
 models for language understanding, and has been applied to a few business domains in our company. 

\textbf{Multi-Interests}. Multi-interest model \cite{cov2016youtube,weston2013interests}  based recommender systems are widely used in our service platform, to capture users' various interests. 

\textbf{GCN}. GCN (Graph Convolutional neural Network) \cite{wang2018billion,ying2018graph} is based on a well-known graph embedding framework. The item embeddings are employed to compute pairwise similarities between all items to facilitate recommendation. 


\begin{table}[!htbp]
\scriptsize
\vspace{-0.3cm}
\caption{Model Scale}
\label{model_config}
\vspace{-0.2cm}
\centering
\begin{tabular}{ c|c|c|c|c}
\hline
&Domain &Dense  &Embedding  & System  \\
& &weights &weights &  Architecture \\
\hline
ResNet50 & CV &204MB &0MB & \emph{AllReduce-Local} \\
\hline
NMT &Translation &706MB &819MB & \emph{AllReduce-Local}  \\
\hline
BERT &QA &1GB &284MB &  \emph{AllReduce-Local}  \\
\hline
Speech &Speech recognition &416MB &0MB &  \emph{1w1g}  \\
\hline
Multi-Interests &Recommender &1.19MB &239.45GB &  \emph{PS/Worker}  \\
\hline
GCN &Recommender &207MB &54GB & \emph{PEARL}\footnotemark \\
\hline
\end{tabular}
\vspace{-0.3cm}
\end{table}

Table \ref{model_config} summarizes the parameter sizes for the models, including dense weights and embedding weights \cite{wang2018billion}. Note that the parameter sizes include both the trainable variables and the optimization-related variables, such as momentums \cite{ruder2016overview}.
For models with small weight size (such as ResNet50, NMT and BERT),
all parameters can concurrently reside in the GPU memory;
hence \emph{AllReduce-Local} architecture is adopted for their training, to leverage GPU-direct technology (NVLink). For models with large-scale weights (such as \emph{Multi-Interests}),
only the \emph{PS/Worker} architecture is suitable, as the weight size supported by the current \emph{AllReduce} frameworks is limited by single GPU's memory size.

In our testbed, we train each model 
using the system  architecture indicated in Table \ref{model_config}. 
The Speech model evaluated is only trained on a small dataset, so does not require distributed training and is trained using \emph{1w1g}. For GCN with a large model size, we will show 
that the limited Ethernet bandwidth becomes the bottleneck when PS/worker architecture is used, and we will design a new system architecture (PEARL) for its training. 
Table \ref{model_character} shows the basic workload features.

\begin{table}[!htbp]
\scriptsize
\vspace{-0.3cm}
\caption{Basic Workload Features}
\label{model_character}
\centering
\begin{tabular}{ p{48pt}| p{17pt} | p{26pt}|p{30pt}|p{37pt} | p{25pt}}
\hline
& Batch Size& FLOP count &Memory access &Memory Copy(PCIe) &Network Traffic \\
\hline
Multi-Interests & 2048 &105.8G &100.4GB &261MB &122MB \\
\hline
ResNet50 & 64 &1.56T &31.9GB &38MB &357MB \\
\hline
NMT & 6144 &2.5T &101.6GB &22KB &1.33GB \\
\hline
BERT & 12 &2.1T &107.3GB &46KB &1.5GB \\
\hline
Speech & 32 &7.9T &20.4GB &804MB &728MB \\
\hline
GCN & 512 & 330.7G &25.79GB &1.2MB &3GB \\
\hline
\end{tabular}
\vspace{-0.4cm}
\end{table}

\subsection{Model Validation}

We first compare the execution time breakdown estimated using the analytical models in Sec. \ref{sec_workflow} and the actual measurement results.
For example, ResNet50 involves 1.56T FLOPs,
while the peak computing FLOPs provided by Tesla V100 in our testbed is 15 TFLOPs;
thus, the compute-bound computation time is predicted via $\frac{1.56}{15*70\%}=0.149s$,
where 70\% is the basic assumption for hardware utilization efficiency.
The actual measured time for this part is 0.126s.
Similar estimation method is used to other parts, including data I/O, weight/gradient traffic time, \emph{etc}.
The estimated time and the actual measurement time, and even the time composition, are used in comparison for model validation.

\begin{figure}[!htb]
\vspace{-0.3cm}
\centering
\centerline{\includegraphics[width=0.9\linewidth]{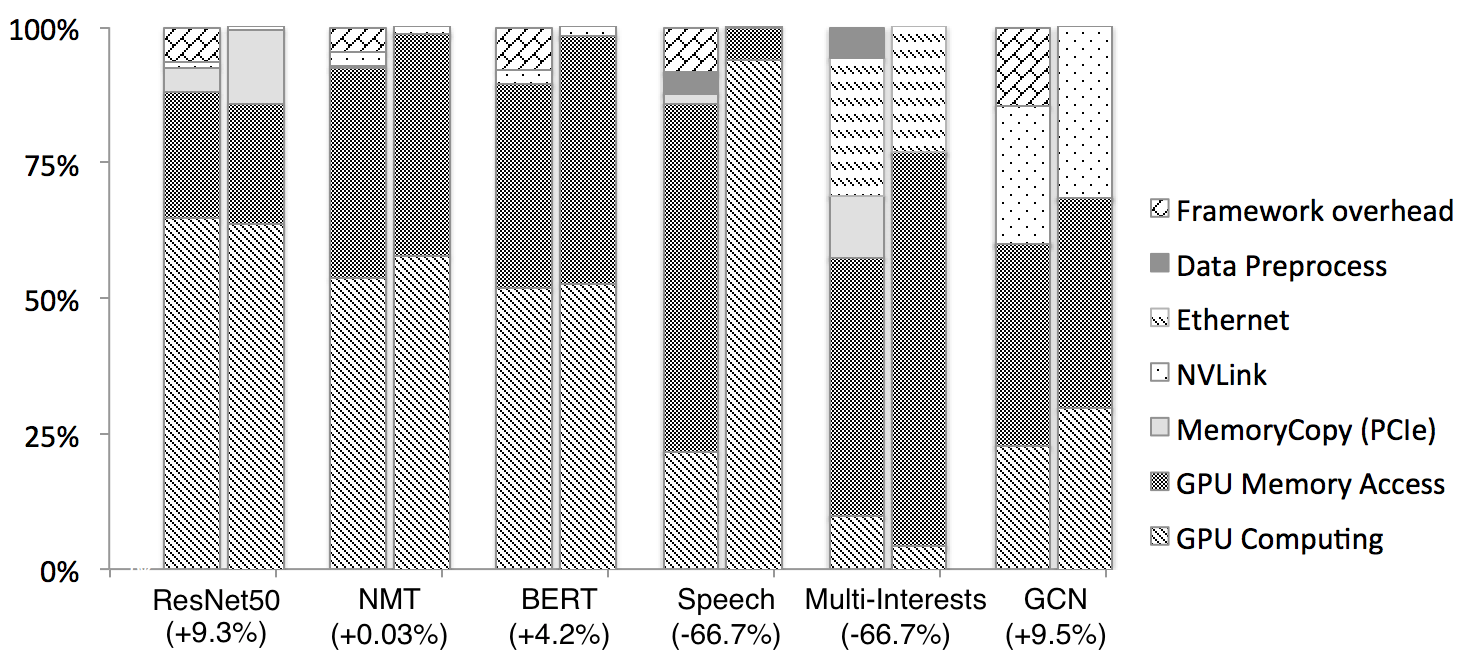}}
\vspace{-0.3cm}
\caption{Time Breakdown Comparison. Left: actual measurement, right: estimation.}
\label{fig_breakdown_comparison}
\end{figure}

\begin{figure*}[!htb]
\vspace{-0.3cm}
\centering
	\centerline{\includegraphics[width=\linewidth]{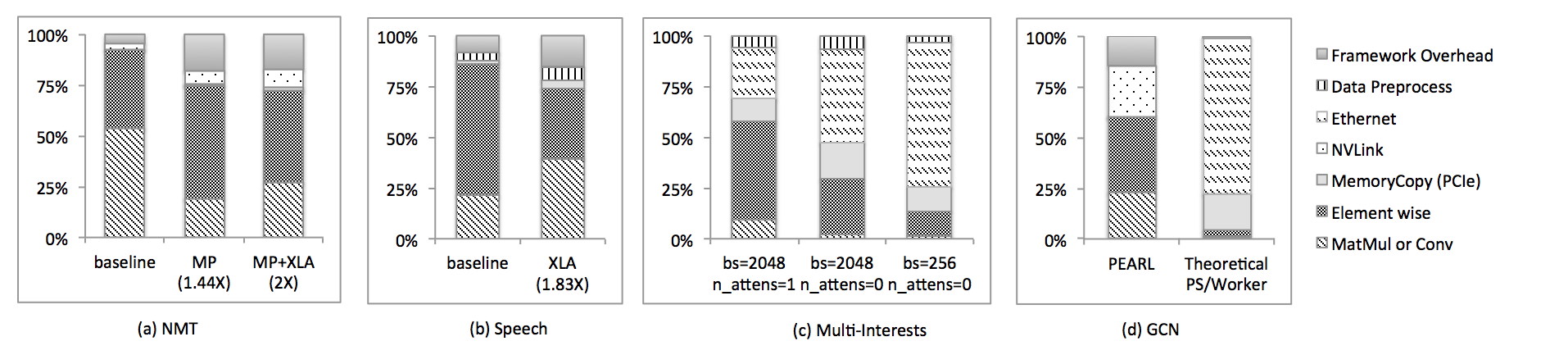}}
	\vspace{-0.5cm}
 	\caption{Performance Breakdown with Different Optimization Techniques.}
 	\label{fig_case_opt}
 	\vspace{-0.5cm}
\end{figure*} 

In Fig. \ref{fig_breakdown_comparison},
the percentage in the parentheses indicates the time difference, computed as $\frac{T_{predict}-T_{actual}}{T_{actual}}$, where $T_{predict}$ is the total time we estimated and $T_{actual}$ is the actual measured time.
The difference is less than 10\% in most cases, and the estimated time breakdown can quite accurately reflect the relative portions of computation and data transfer in the entire execution time.
For the Speech model, the difference is more than 66.7\%.
Our estimation inaccuracy is due to the actual low usage of GPU memory access bandwidth at only 3\%, 
much smaller than the 70\% assumption when we do the estimation. 
We seek how to further improve memory access efficiency as a future direction, while adopting possible optimization such as XLA to reduce the memory-access volume by operation fusion, to accelerate training of the Speech model.

\vspace{-0.2cm}

\subsection{PEARL Architecture}
\label{sec_pearl_arch}
 
Used in the domain of e-commerce, search and recommendation models have very large and sparse commodity-embedding parameters. When the model size (ranging from tens to hundreds of GB) is too large to fit into the GPU memory entirely, the \emph{PS/Worker} architecture should be adopted to partition and store the variables in the CPU memory among multiple \emph{PS} nodes. However, synchronizing a large variable among the PS and GPUs of the workers requires significant ethernet and PCIe bandwidth, and also consumes many CPU clocks.

Parameters of such models can be classified into dense and sparse weights, depending on how their elements are accessed. Treating the whole model as dense is inefficient, since naïvely communicating all elements of a large sparse variable, even though only a small subset is accessed, results in relatively low scalability.

We propose and implement PEARL (Partitioned Embedding And RepLicated), a new distribution strategy that optimizes the efficiency of data transfer by taking the sparsity of variables into account.

\begin{figure}[!htb]
\vspace{-0.1cm}
\centering
	\centerline{\includegraphics[width=0.9\linewidth]{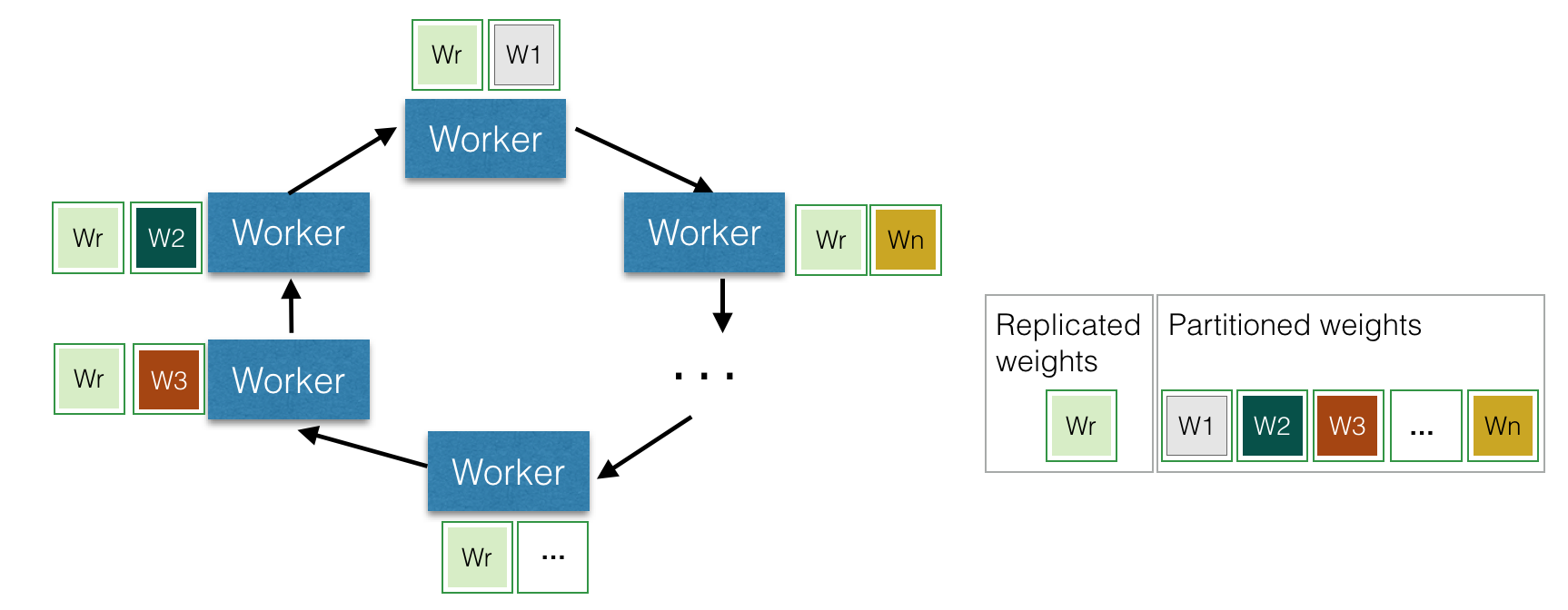}}
	\vspace{-0.3cm}
 	\caption{Architecture of PEARL.}
 	\label{fig_PEARL}
 	\vspace{-0.1cm}
\end{figure}

As shown in Fig.~\ref{fig_PEARL}, PEARL applies a hybrid approach that partitions the large sparse variables and distributes them in the GPU memory of workers, and adopts the AllReduce architecture to process dense variables. 

All workers synchronize variables via collective communication operations such as AllReduce and AllGatherv. AllReduce aggregates gradients from all GPUs for the dense weights, while AllGatherv gathers the embedding weights and corresponding gradients from all GPUs for the partitioned weights. The AllGatherv operation is implemented on top of NCCL \cite{nccl2018} primitives such as Broadcast and Reduce, that are optimized to leverage high-speed inter-GPU NVLink. 

Experiments show that PEARL built atop TensorFlow achieves good scalability in terms of training throughput with the increase of computation resources, on both dense and sparse models.

\subsection{Effectiveness of Optimization Techniques}


As shown in Fig. \ref{fig_breakdown_comparison}, behavior of ResNet50, NMT and BERT is
quite similar:
1) the actual time measurements and the model-based estimation are close,
indicating that the hardware usage efficiency is quite high, around the basic assumption of 70\%.
2) the computation part contributes the most to the total running time, which shows that the communication part of time is reduced quite well by using NVLink. 

We next investigate how to further improve the computation efficiency. 
Fig. \ref{fig_case_opt}(a) compares the results obtained using the default setting, 
with mixed-precision (MP) matrix multiplication in FP16 \cite{micikevicius2017mixed} enabled (which is available with TensorCore in Volta architecture, potentially achieving up to 8X speedup compared to the default multiply-and-addition in FP32), and with XLA enforced. 
We observe 1.44X end-to-end speedup and 2.8X for MatMul when mixed-precision optimization is in use. With the powerful tool XLA (operation fusion and code generation), element-wise operation time can be reduced, as operation fusion exploits GPU's high-speed cache to reduce the framework scheduling overhead. We observe 2X speedup with both MP and XLA in place (1.76X with only XLA).





Fig. \ref{fig_case_opt}(b) shows that when using XLA when training the Speech model,
3.43X speedup can be achieved for element-wise operations and 1.83X for the end-to-end performance.


Figure \ref{fig_case_opt}(c) presents the time breakdown of Multi-Interests model training under three different training configurations (batch size and the number of attention layers). With the same model, performance bottlenecks in case of different configurations vary significantly. Larger batch size is more friendly to GPU with element-wise operations being the bottleneck, whose computation time can be reduced by operation fusion at the runtime. With the third configuration, communication becomes the bottleneck.
A Multi-Interests model has a large weight size of more than 200GB; the weights cannot be entirely stored in the GPU memory. 
Therefore, we cannot apply the \emph{AllReduce} architecture to leverage the high-speed NVLink (since current AllReduce frameworks only support weight-replica mode).
Similarly, GCN has large-scale embedding weights, and \emph{PS/Worker} framework should be used.
However, large-volume communication via Ethernet and PCIe may become the bottleneck. In these cases, 
PEARL is applied, which can use NVLink to transfer the weights/gradients for large-scale models, is in need.

With PEARL, the large-scale weights, such as embeddings, are partitioned among multiple GPUs, while the variable/gradient 
aggregation is performed using a PS/Worker like protocol, using \emph{AllGather} and \emph{ReduceScatter} operations \cite{nccl2018};
all other small-scale weights
are replicated and \emph{AllReduce} is used for gradient exchange. 
Fig. \ref{fig_case_opt}(d) presents the time breakdown of GCN model when trained using PEARL. We see that with the high-speed interconnect, the communication part via NVLink consumes 25\%
of the total time. Using our analytical approach, we can also estimate the time breakdown when using \emph{PS/Worker} with Ethernet \& PCIe for training, which is shown in the second bar in Fig. \ref{fig_case_opt}(d).
The communication part with Ethernet \& PCIe contributes to almost 95\% of the total time, which is much more than what we can achieve with PEARL. 


\section{Discussions}
In our proposed workload characterization framework, there are several assumptions that may affect the results.
In this section, we discuss the effects when the assumptions shift.

\subsection{Hardware efficiency assumption}
As described in Sec. \ref{sec_workflow}, 
the hardware efficiencies of computation (GPU) and communication (PCIe/Ethernet/NVLink) parts are both assumed to be 70\%.
To find out whether the assumption is reasonable, we conduct cross-validation in two ways. 
First, we measure the hardware efficiency in each case analyzed in Sec. \ref{sec_casestudy}. 
Next, as it is sophisticated to establish a system to precisely measure the hardware utility efficiency for each workload,
instead we try to analyze how the results will shift if the assumption is not followed.

Table \ref{table_model_efficiency} shows the actual measured hardware efficiency for each workload.
70\% is about the average level. 
In detail,
we can observe that, the efficiency of GPU computation/ memory access is a bit higher than 70\%, while that of data traffic (PCIe/Ethernet/NVlink) is lower.

\begin{table}[!htbp]
\caption{Resource Efficiency for Each Workload}
\label{table_model_efficiency}
\centering
\begin{tabular}{ p{50pt}|p{30pt}|p{32pt}|p{28pt}|p{35pt}}
\hline
&GPU TOPS &GDDR &PCIe &Network (Ethernet/NVLink) \\
\hline
Multi-Interests &32.71\% &95\% &86.47\% &69.21\% \\
\hline
ResNet50 &82.55\% &78.9\% &35.1\% &49.4\% \\
\hline
NMT &82.8\% &79.1\% &0.1\% &35.2\%\\
\hline
BERT &81.6\% &95\% &0.42\% &47.1\% \\
\hline
Audio &60.86\% &3.1\% &77.73\% &40.5\%\\
\hline
GCN &88.2\% &69.9\% &86.2\% &27.35\% \\
\hline
\end{tabular}
\end{table}

\begin{figure}[!htb]
\vspace{-0.3cm}
\centering
	\centerline{\includegraphics[width=0.8\linewidth]{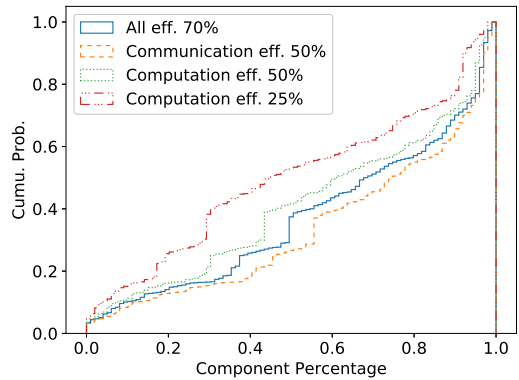}}
	\caption{Shift Effect in Weight Traffic Percentage When Hardware Efficiency Changes.}
	\label{fig_hardware_eff}
\end{figure}

For the collective behavior, we explore how the conclusion will change if the assumption is violated.
Taking \emph{PS/Worker} workloads as example,
we evaluate how the weight traffic's portion in the end-to-end training time varies when the hardware efficiency in computation/communication changes.
As shown in Fig. \ref{fig_hardware_eff},
as expected, when the actual hardware efficiency in communication (PCIe/Ethernet) is lower than 70\%, the \emph{PS/Worker} workloads will spend more time on weight traffic,
and vice versa.
What should be noted is that even when the hardware efficiency in computation is only 25\% (quite lower than the 70\% assumption), the \emph{PS/Worker} workloads still spend more time on weight traffic on average.

To give a precise estimation on the fundamental bottleneck in the cluster using our proposed framework, it is still important to establish a better methodology to measure the utilization efficiency of each hardware component, which will be an important direction in our future work.

\subsection{Computation/communication overlap assumption}
There are various ways to overlap computation and data transfer \cite{zhang2017poseidon,hashemi2018tictac} in DL workloads.
Although the purpose of this work is to expose the fundamental performance bottlenecks,
which will not change due to the overlap issue,
several speedup results may change if the non-overlap assumption is violated.
As how to achieve computation and communication overlap is still an open question in deep learning design,
it is not easy to quantify the actual overlap potential for each workload.
Instead, we use an ideal overlap case to give an estimation for comparison. In this case, the total time changes from $T_{total}=T_d + T_c + T_w$ (used in our framework in Sec. \ref{sec_workflow}) to $T_{total} = \max\{T_d, T_c, T_w\}$.

\begin{figure}[!htb]
\vspace{-0.3cm}
\centering
	\begin{minipage}[b]{0.48\linewidth}
	\centerline{\includegraphics[width=\linewidth]{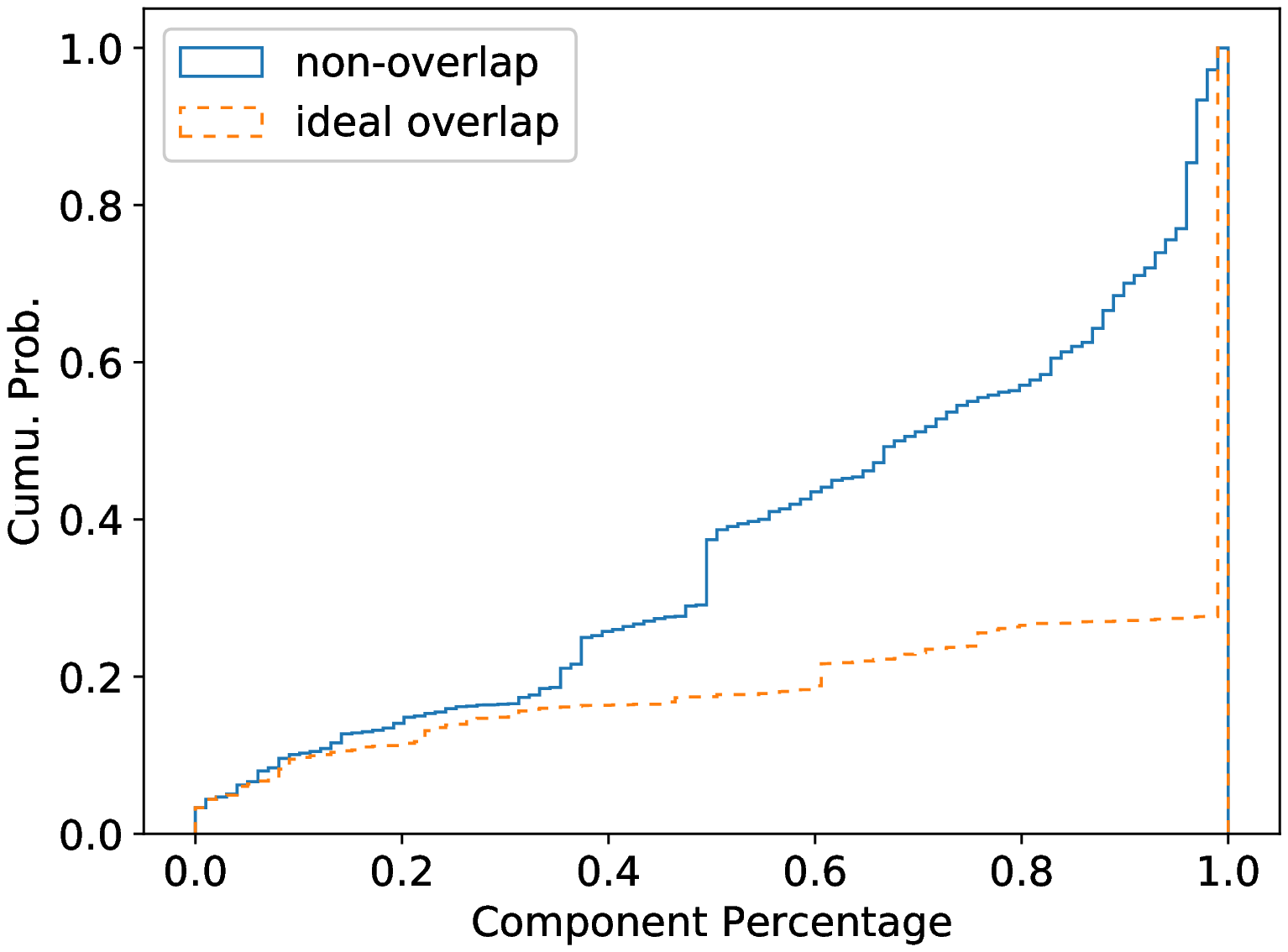}}
	\end{minipage}
	\begin{minipage}[b]{0.48\linewidth}
	\centerline{\includegraphics[width=\linewidth]{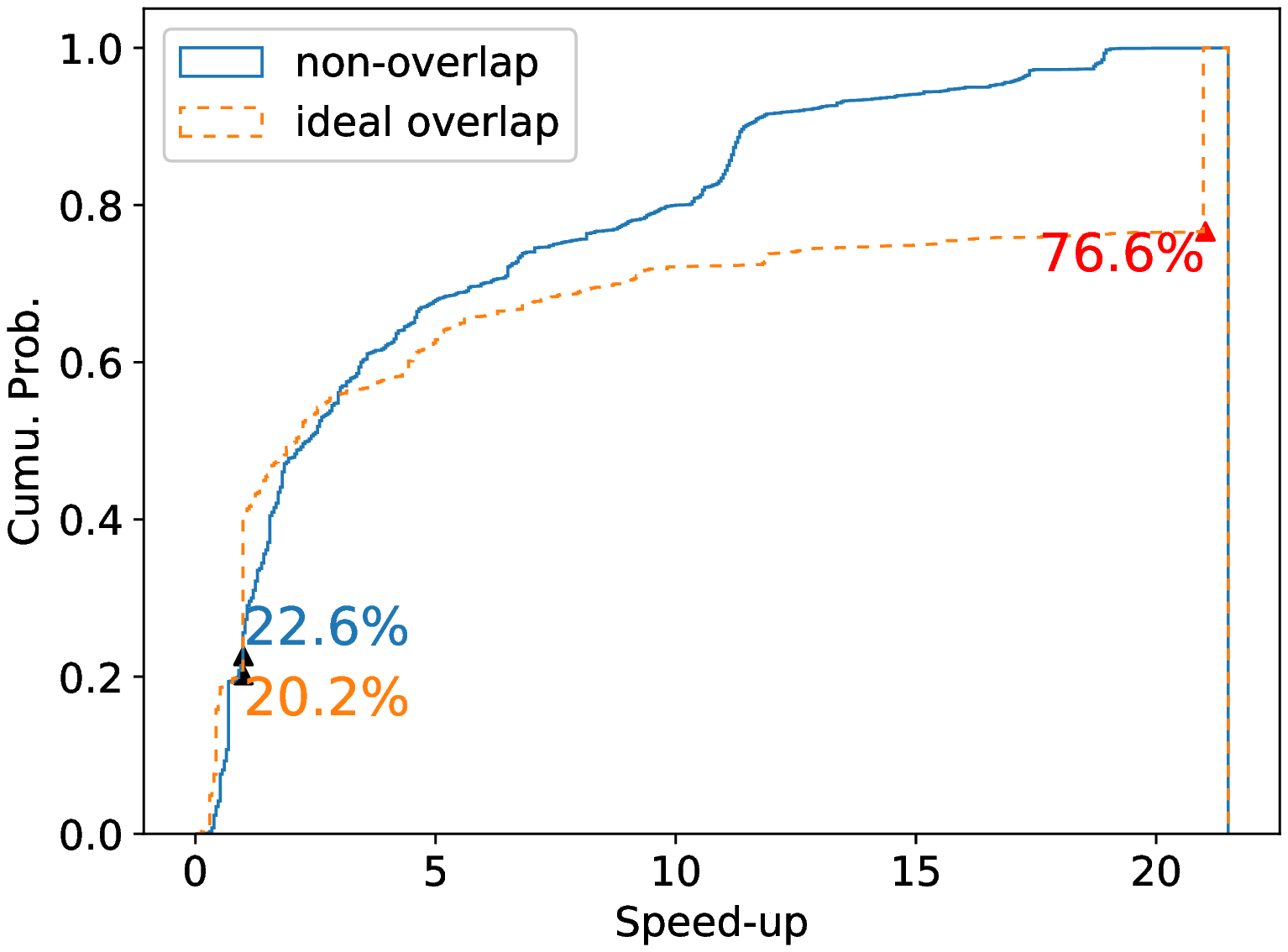}}
	\end{minipage}
	\caption{Shift Effect Under Different Overlap States. Left: weight traffic percent, right: speedup when mapping to \emph{AllReduce-Local}}
	\label{fig_overlap}
\end{figure}

Fig. \ref{fig_overlap} shows the comparison results of \emph{PS/Worker} workloads under different overlap states: totally none-overlap \emph{VS} ideal-overlap.
It can be observed that when computation and communication ideally overlap, the weight traffic part is heavily exposed as the performance bottleneck,
as it consumes the longest time among $\{T_d, T_c, T_w\}$. As to the speedup analysis when mapping \emph{PS/Worker} workloads to the \emph{AllReduce-Local} architecture,
we can observe that the ratio of sped-up workloads remains similar as the none-overlap results, 22.6\% \emph{VS} 20.2\%.
It can be noted that there are 23.4\% workloads achieving 21X speedup, which are actually the workloads bound by the weight traffic part either before or after the architecture projection. For such workloads, the speedup ratio can be computed as:
\begin{equation}
\frac{\frac{S_w}{25Gb\times 70\%} + \frac{S_w}{10GB\times 70\%}}{\frac{S_w}{10GB\times 70\%}}=21
\end{equation}
where $S_w$ denotes the weight traffic volume.

The comparison further illustrates that the assumption of computation/communication overlap may affect the detailed analytical results, such as speedup ratio or running time constitution; however, it does not change the conclusion as to what is the fundamental bottleneck for the workloads in our cluster.
At last, it is worth noting that the purpose of our analysis framework is not to precisely predict practical performance of workloads, but to expose the fundamental bottlenecks in hardware components or system architecture for collective behavior of workloads in our cluster.

\section{System Implications}
\label{sec_implication}
Based on previous results, we now summarize important
implications on how to optimize training frameworks (\emph{e.g.} TensorFlow) and
 how to properly provision system resources.
 
\subsection{Implications on Framework Optimization}

\subsubsection{System Architecture}
In the PAI cluster, we identified plenty of DL models that are not suitable to be trained using either \emph{PS/Worker} or \emph{AllReduce}, \emph{e.g.}, models with one large sparse embedding and many 
relatively small dense weights (such as GCN in Section \ref{sec_casestudy}). 
The weight sizes within such workloads are too large to be resident in GPU memory.
On the other hand, such workloads always incur heavy weight/gradient traffic, for which the Ethernet connections with limited bandwidth will be the bottleneck.
For such workloads, we proposed a new strategy PEARL, as inspired by our characterization of collective behavior in the cluster, catering to the resource requirements of such workloads.

Our simple analytical model can predict the time breakdown of jobs on different architectures, facilitating system architecture selection. Though our model does not take potential framework overhead into consideration, experiments show that its estimation is quite close to real measurements for representative workloads.
A more comprehensive prediction method 
 is one of our future directions to explore.

\subsubsection{Compilation: Operation Fusion and Code Generation}
Statistical results in Sec. \ref{sec_clusterbehavior} show that, within the computation part,
the time spent on memory-bound operations is no less than that of computation-bound ones.
TensorFlow XLA is a solid \mengdi{compilation} framework for operation fusion and code generation
to reduce the memory accesses.
We have shown that XLA is powerful enough to handle practical training workloads. 
As shown in Sec. \ref{sec_casestudy},
different workloads have drastically different computation profiles.
For ResNet50, NMT and BERT, memory-access time takes at most
40\% of execution time. In large-scale recommendation models (Multi-Interests,
GCN), it takes up to 60\%. For all these workloads, compilation using XLA is helpful in
reducing CPU launch overhead and improving GPU computation efficiency.

XLA is known to have several limitations.
For example, it cannot deal well with workloads with dynamic shapes,
the operation fusion algorithm is designed as rule-based and cannot be generalized well,
the code generation mechanism still needs to be improved to generate highly optimized 
kernels
\cite{FusionStitching}.
The community is calling for a powerful, 
robust compilation infrastructure that
is able to handle rapidly changing training
workloads in the future.

\subsubsection{Framework Overhead}
Frameworks like TensorFlow use a flexible and general enough CPU runtime to do
computation scheduling. If the main part of
the computation graph consists of very fine-grained operations, CPU scheduling may 
incur non-negligible overheads, especially in busy CPU/GPU clusters with a mixture
of workloads deployed.

Most of our workloads have regular computation structures, and carry out repetitive iterations during the  training process. Through compilation (discussed above),
it is possible to allow a larger portion of the computation graph to be scheduled to the GPU altogether. 

\subsection{Implications on Hardware Configurations}

\subsubsection{Interconnect Bandwidth}
There are two types of interconnects for distributed training in our cluster:
NVLink and Ethernet, with notable gap \emph{w.r.t.} the communication bandwidth.
We have shown the performance gain
of high speed interconnects for weight/gradient communication in numerous medium scale ($<$50GB) workloads. For large models
(\emph{e.g.} \emph{Multi-Interests} model in Sec. \ref{sec_casestudy}), 
weight/gradient communication over the Ethernet can take up to more than 
50\% of execution time per iteration. High-bandwidth interconnects will definitely
help such communication-bound workloads, as shown in 
 Fig. \ref{fig_resource_provision}.

\subsubsection{PCIe Bandwidth}
In our system settings, PCIe is mainly dedicated for data transfer between 
CPU and GPU. In distributed training, PCIe traffic normally
consists of two portions: sample data input, and weight/gradient communication.
As shown in Sec. \ref{sec_clusterbehavior}, in most workloads, the sample 
input volume is negligible, and weight/gradient transfer is usually bound by 
network rather than PCIe.

However, this does not mean that PCIe bandwidth is less important for performance.
As shown in Fig. \ref{fig_breakdown_allreduce}, the bottleneck may be shifted to PCIe after the network bandwidth usage is optimized.
Additionally, high-speed PCIe interconnects can enable exciting new
optimization opportunities for some mission critical applications. The basic idea is to push
as much work as possible, from CPU to GPU, in order to allow more operations in the computation graph to be processed in GPU as a whole
and minimize 
CPU intervention.

\subsubsection{GPU Computing Power and Memory Bandwidth}
Computing power and memory bandwidth of GPUs are essential for DL workloads. 
Important as they are, we have shown in Sec. \ref{sec_clusterbehavior} that weight/gradient communication renders the biggest performance bottleneck in our cluster.
More careful
model distribution and system architecture selection are necessary to mitigate 
communication overhead in order to fully exploit the computation power.

\section{Related Work}
\label{sec_related}

There have recently been several studies conducting cluster-level machine learning workload characterization,
aiming to improve resource utilization and workload performance in the ML cluster \cite{park2018deep,jeon2018multi,cortez2017resource}.
Park \emph{et al.} \cite{park2018deep} analyze the inference workloads in a Facebook data center,
pointing out limitations of the current ML infrastructure \cite{hazelwood2018applied}
and providing suggestions for future general-purpose/accelerated inference hardware.
Jeon \emph{et al.} \cite{jeon2018multi} present a detailed workload characterization of a two-month trace
from a multi-tenant GPU cluster, and focused on resource utilization and scheduling.

Some other work aim to establish the performance benchmark \cite{zhu2018benchmarking,adolf2016fathom,gao2018data,gao2018data2}. 
Fathom \cite{adolf2016fathom} establishes a set of reference implementation for 
eight archetypal DL jobs.
Guignard \emph{et al.} \cite{guignard2018performance} adopt the eight types of workloads from Fathom to evaluate the performance of the IBM ``Minsky'' platform. 
A micro-benchmark is designed in \cite{chien2018characterizing} to measure reads in TensorFlow and a burst buffer is
implemented to improve the I/O performance.
Gao \emph{et al.} \cite{gao2018data,gao2018data2} establish a proxy benchmark for AI workloads by identifying eight data motifs. 

Several studies have focused on predicting the performance of
a job using a mathematical model \cite{qi2016paleo,gu2017deepprof,venkataraman2016ernest,bakhoda2009analyzing}. PALEO \cite{qi2016paleo} establishes a performance
model by extracting the basic computational requirements and mapping them to a specific point
within the design space of software, hardware and communication strategies.
DeepProf \cite{gu2017deepprof} is a tool that can automatically process GPU traces
and generate performance reports for deep learning applications,
which can perform diagnosis to identify the runtime bottleneck. 
The above two work both aim to break down the execution time of a workload,
with the former analyzing from the theoretical perspective 
and the latter using runtime traces.
Ernest \cite{venkataraman2016ernest} builds a performance model from the workload observation
on small datasets and predicts the performance on larger datasets in bigger clusters.
Justus \emph{et al.} \cite{justus2018predicting} predict execution time of one part in the entire DL network; execution time of the sub-graph constitutes a basic unit for predicting the end-to-end performance.

Different from the existing work that aim at 
precisely predicting practical performance of a given workload,
\mengdi{our work focuses on characterization of currently deployed jobs on our large cluster and extracting their fundamental resource requirements,
in order to expose the potential hardware/software optimization directions at the cluster scale.}
From our observations, we extract fundamental execution 
bottlenecks and identify latent, useful directions for training framework optimization or system configuration improvement. 

\section{Conclusion}
\label{sec_conclusion}
This paper presents a characterization framework to enable performance 
analysis over diversified production workloads running on Alibaba-PAI.
The framework features a lightweight technique to collect runtime profiling metrics 
of workloads. Based on collected job statistics, we build a workload model
to extract key features and project them to different system configurations in order to analytically predict the performance behavior. 
We characterize collective behavior of a large volume of workloads, as well as zoom into representative workloads for investigating impact of different system architectures and hardware configurations. 
We discuss potential technical directions for improving training performance of the workloads.
As future work, we seek to characterize inference workloads in our cluster using a similar methodology. 

\bibliographystyle{unsrt}
\bibliography{workload}

\begin{thebibliography}{10}

\bibitem{krizhevsky2012imagenet}
Alex Krizhevsky, Ilya Sutskever, and Geoffrey~E Hinton.
\newblock Imagenet classification with deep convolutional neural networks.
\newblock In {\em Advances in neural information processing systems}, pages
  1097--1105, 2012.

\bibitem{he2016identity}
Kaiming He, Xiangyu Zhang, Shaoqing Ren, and Jian Sun.
\newblock Identity mappings in deep residual networks.
\newblock In {\em European conference on computer vision}, pages 630--645.
  Springer, 2016.

\bibitem{karpathy2014large}
Andrej Karpathy, George Toderici, Sanketh Shetty, Thomas Leung, Rahul
  Sukthankar, and Li~Fei-Fei.
\newblock Large-scale video classification with convolutional neural networks.
\newblock In {\em Proceedings of the IEEE conference on Computer Vision and
  Pattern Recognition}, pages 1725--1732, 2014.

\bibitem{simonyan2014two}
Karen Simonyan and Andrew Zisserman.
\newblock Two-stream convolutional networks for action recognition in videos.
\newblock In {\em Advances in neural information processing systems}, pages
  568--576, 2014.

\bibitem{bahdanau2014neural}
Dzmitry Bahdanau, Kyunghyun Cho, and Yoshua Bengio.
\newblock Neural machine translation by jointly learning to align and
  translate.
\newblock {\em arXiv preprint arXiv:1409.0473}, 2014.

\bibitem{devlin2018bert}
Jacob Devlin, Ming-Wei Chang, Kenton Lee, and Kristina Toutanova.
\newblock Bert: Pre-training of deep bidirectional transformers for language
  understanding.
\newblock {\em arXiv preprint arXiv:1810.04805}, 2018.

\bibitem{graves2013speech}
Alex Graves, Abdel-rahman Mohamed, and Geoffrey Hinton.
\newblock Speech recognition with deep recurrent neural networks.
\newblock In {\em 2013 IEEE international conference on acoustics, speech and
  signal processing}, pages 6645--6649. IEEE, 2013.

\bibitem{chorowski2015attention}
Jan~K Chorowski, Dzmitry Bahdanau, Dmitriy Serdyuk, Kyunghyun Cho, and Yoshua
  Bengio.
\newblock Attention-based models for speech recognition.
\newblock In {\em Advances in neural information processing systems}, pages
  577--585, 2015.

\bibitem{wang2018billion}
Jizhe Wang, Pipei Huang, Huan Zhao, Zhibo Zhang, Binqiang Zhao, and Dik~Lun
  Lee.
\newblock Billion-scale commodity embedding for e-commerce recommendation in
  alibaba.
\newblock In {\em Proceedings of the 24th ACM SIGKDD International Conference
  on Knowledge Discovery \& Data Mining}, pages 839--848. ACM, 2018.

\bibitem{ying2018graph}
Rex Ying, Ruining He, Kaifeng Chen, Pong Eksombatchai, William~L Hamilton, and
  Jure Leskovec.
\newblock Graph convolutional neural networks for web-scale recommender
  systems.
\newblock In {\em Proceedings of the 24th ACM SIGKDD International Conference
  on Knowledge Discovery \& Data Mining}, pages 974--983. ACM, 2018.

\bibitem{chen2015deepdriving}
Chenyi Chen, Ari Seff, Alain Kornhauser, and Jianxiong Xiao.
\newblock Deepdriving: Learning affordance for direct perception in autonomous
  driving.
\newblock In {\em Proceedings of the IEEE International Conference on Computer
  Vision}, pages 2722--2730, 2015.

\bibitem{silver2016mastering}
David Silver, Aja Huang, Chris~J Maddison, Arthur Guez, Laurent Sifre, George
  Van Den~Driessche, Julian Schrittwieser, Ioannis Antonoglou, Veda
  Panneershelvam, Marc Lanctot, et~al.
\newblock Mastering the game of go with deep neural networks and tree search.
\newblock {\em nature}, 529(7587):484, 2016.

\bibitem{zoph2016neural}
Barret Zoph and Quoc~V Le.
\newblock Neural architecture search with reinforcement learning.
\newblock {\em arXiv preprint arXiv:1611.01578}, 2016.

\bibitem{shi2016benchmarking}
Shaohuai Shi, Qiang Wang, Pengfei Xu, and Xiaowen Chu.
\newblock Benchmarking state-of-the-art deep learning software tools.
\newblock In {\em 2016 7th International Conference on Cloud Computing and Big
  Data (CCBD)}, pages 99--104. IEEE, 2016.

\bibitem{gu2017deepprof}
Jiazhen Gu, Huan Liu, Yangfan Zhou, and Xin Wang.
\newblock Deepprof: Performance analysis for deep learning applications via
  mining gpu execution patterns.
\newblock {\em arXiv preprint arXiv:1707.03750}, 2017.

\bibitem{adolf2016fathom}
Robert Adolf, Saketh Rama, Brandon Reagen, Gu-Yeon Wei, and David Brooks.
\newblock Fathom: Reference workloads for modern deep learning methods.
\newblock In {\em 2016 IEEE International Symposium on Workload
  Characterization (IISWC)}, pages 1--10. IEEE, 2016.

\bibitem{li2018tartan}
Ang Li, Shuaiwen~Leon Song, Jieyang Chen, Xu~Liu, Nathan Tallent, and Kevin
  Barker.
\newblock Tartan: Evaluating modern gpu interconnect via a multi-gpu benchmark
  suite.
\newblock In {\em 2018 IEEE International Symposium on Workload
  Characterization (IISWC)}, pages 191--202. IEEE, 2018.

\bibitem{gao2018data2}
Wanling Gao, Jianfeng Zhan, Lei Wang, Chunjie Luo, Zhen Jia, Daoyi Zheng, Chen
  Zheng, Xiwen He, Hainan Ye, Haibin Wang, et~al.
\newblock Data motif-based proxy benchmarks for big data and ai workloads.
\newblock In {\em 2018 IEEE International Symposium on Workload
  Characterization (IISWC)}, pages 48--58. IEEE, 2018.

\bibitem{hazelwood2018applied}
Kim Hazelwood, Sarah Bird, David Brooks, Soumith Chintala, Utku Diril, Dmytro
  Dzhulgakov, Mohamed Fawzy, Bill Jia, Yangqing Jia, Aditya Kalro, et~al.
\newblock Applied machine learning at facebook: a datacenter infrastructure
  perspective.
\newblock In {\em 2018 IEEE International Symposium on High Performance
  Computer Architecture (HPCA)}, pages 620--629. IEEE, 2018.

\bibitem{park2018deep}
Jongsoo Park, Maxim Naumov, Protonu Basu, Summer Deng, Aravind Kalaiah, Daya
  Khudia, James Law, Parth Malani, Andrey Malevich, Satish Nadathur, et~al.
\newblock Deep learning inference in facebook data centers: Characterization,
  performance optimizations and hardware implications.
\newblock {\em arXiv preprint arXiv:1811.09886}, 2018.

\bibitem{shi2018performance}
Shaohuai Shi, Qiang Wang, and Xiaowen Chu.
\newblock Performance modeling and evaluation of distributed deep learning
  frameworks on gpus.
\newblock In {\em 2018 IEEE 16th Intl Conf on Dependable, Autonomic and Secure
  Computing, 16th Intl Conf on Pervasive Intelligence and Computing, 4th Intl
  Conf on Big Data Intelligence and Computing and Cyber Science and Technology
  Congress (DASC/PiCom/DataCom/CyberSciTech)}, pages 949--957. IEEE, 2018.

\bibitem{zhu2018benchmarking}
Hongyu Zhu, Mohamed Akrout, Bojian Zheng, Andrew Pelegris, Anand Jayarajan,
  Amar Phanishayee, Bianca Schroeder, and Gennady Pekhimenko.
\newblock Benchmarking and analyzing deep neural network training.
\newblock In {\em 2018 IEEE International Symposium on Workload
  Characterization (IISWC)}, pages 88--100. IEEE, 2018.

\bibitem{volta}
NVIDIA.
\newblock Nvidia tesla v100 gpu architecture.
\newblock
  \url{https://images.nvidia.com/content/volta-architecture/pdf/volta-architecture-whitepaper.pdf},
  2017.

\bibitem{XLA}
The~XLA Team.
\newblock Xla – tensorflow compiled. post in the google developers blog.
\newblock
  \url{https://developers.googleblog.com/2017/03/xla-tensorflow-compiled.html},
  2017.

\bibitem{abadi2016tensorflow}
Mart{\'\i}n Abadi, Paul Barham, Jianmin Chen, Zhifeng Chen, Andy Davis, Jeffrey
  Dean, Matthieu Devin, Sanjay Ghemawat, Geoffrey Irving, Michael Isard, et~al.
\newblock Tensorflow: A system for large-scale machine learning.
\newblock In {\em 12th $\{$USENIX$\}$ Symposium on Operating Systems Design and
  Implementation ($\{$OSDI$\}$ 16)}, pages 265--283, 2016.

\bibitem{jia2014caffe}
Yangqing Jia, Evan Shelhamer, Jeff Donahue, Sergey Karayev, Jonathan Long, Ross
  Girshick, Sergio Guadarrama, and Trevor Darrell.
\newblock Caffe: Convolutional architecture for fast feature embedding.
\newblock In {\em Proceedings of the 22nd ACM international conference on
  Multimedia}, pages 675--678. ACM, 2014.

\bibitem{paszke2017automatic}
Adam Paszke, Sam Gross, Soumith Chintala, Gregory Chanan, Edward Yang, Zachary
  DeVito, Zeming Lin, Alban Desmaison, Luca Antiga, and Adam Lerer.
\newblock Automatic differentiation in pytorch.
\newblock 2017.

\bibitem{chen2015mxnet}
Tianqi Chen, Mu~Li, Yutian Li, Min Lin, Naiyan Wang, Minjie Wang, Tianjun Xiao,
  Bing Xu, Chiyuan Zhang, and Zheng Zhang.
\newblock Mxnet: A flexible and efficient machine learning library for
  heterogeneous distributed systems.
\newblock {\em arXiv preprint arXiv:1512.01274}, 2015.

\bibitem{yu2014introduction}
Dong Yu, Adam Eversole, Mike Seltzer, Kaisheng Yao, Zhiheng Huang, Brian
  Guenter, Oleksii Kuchaiev, Yu~Zhang, Frank Seide, Huaming Wang, et~al.
\newblock An introduction to computational networks and the computational
  network toolkit.
\newblock {\em Microsoft Technical Report MSR-TR-2014--112}, 2014.

\bibitem{NVLink}
Nvlink.
\newblock \url{https://www.nvidia.com/en-gb/data-center/nvlink/}.

\bibitem{mayer2019scalable}
Ruben Mayer and Hans-Arno Jacobsen.
\newblock Scalable deep learning on distributed infrastructures: Challenges,
  techniques and tools.
\newblock {\em arXiv preprint arXiv:1903.11314}, 2019.

\bibitem{shallue2018measuring}
Christopher~J Shallue, Jaehoon Lee, Joe Antognini, Jascha Sohl-Dickstein, Roy
  Frostig, and George~E Dahl.
\newblock Measuring the effects of data parallelism on neural network training.
\newblock {\em arXiv preprint arXiv:1811.03600}, 2018.

\bibitem{li2014scaling}
Mu~Li, David~G Andersen, Jun~Woo Park, Alexander~J Smola, Amr Ahmed, Vanja
  Josifovski, James Long, Eugene~J Shekita, and Bor-Yiing Su.
\newblock Scaling distributed machine learning with the parameter server.
\newblock In {\em 11th $\{$USENIX$\}$ Symposium on Operating Systems Design and
  Implementation ($\{$OSDI$\}$ 14)}, pages 583--598, 2014.

\bibitem{nccl2018}
NVIDIA.
\newblock Nvidia collective communications library.
\newblock \url{https://github.com/NVIDIA/nccl}, May 2018.

\bibitem{goldsborough2016tour}
Peter Goldsborough.
\newblock A tour of tensorflow.
\newblock {\em arXiv preprint arXiv:1610.01178}, 2016.

\bibitem{zhang2017poseidon}
Hao Zhang, Zeyu Zheng, Shizhen Xu, Wei Dai, Qirong Ho, Xiaodan Liang, Zhiting
  Hu, Jinliang Wei, Pengtao Xie, and Eric~P Xing.
\newblock Poseidon: An efficient communication architecture for distributed
  deep learning on $\{$GPU$\}$ clusters.
\newblock In {\em 2017 $\{$USENIX$\}$ Annual Technical Conference
  ($\{$USENIX$\}$$\{$ATC$\}$ 17)}, pages 181--193, 2017.

\bibitem{hashemi2018tictac}
Sayed~Hadi Hashemi, Sangeetha~Abdu Jyothi, and Roy~H Campbell.
\newblock Tictac: Accelerating distributed deep learning with communication
  scheduling.
\newblock {\em arXiv preprint arXiv:1803.03288}, 2018.

\bibitem{tallent2017evaluating}
Nathan~R Tallent, Nitin~A Gawande, Charles Siegel, Abhinav Vishnu, and Adolfy
  Hoisie.
\newblock Evaluating on-node gpu interconnects for deep learning workloads.
\newblock In {\em International Workshop on Performance Modeling, Benchmarking
  and Simulation of High Performance Computer Systems}, pages 3--21. Springer,
  2017.

\bibitem{qi2016paleo}
Hang Qi, Evan~R Sparks, and Ameet Talwalkar.
\newblock Paleo: A performance model for deep neural networks.
\newblock 2016.

\bibitem{dai2016r}
Jifeng Dai, Yi~Li, Kaiming He, and Jian Sun.
\newblock R-fcn: Object detection via region-based fully convolutional
  networks.
\newblock In {\em Advances in neural information processing systems}, pages
  379--387, 2016.

\bibitem{vaswani2017attention}
Ashish Vaswani, Noam Shazeer, Niki Parmar, Jakob Uszkoreit, Llion Jones,
  Aidan~N Gomez, {\L}ukasz Kaiser, and Illia Polosukhin.
\newblock Attention is all you need.
\newblock In {\em Advances in neural information processing systems}, pages
  5998--6008, 2017.

\bibitem{kim2017dynamic}
Taesup Kim, Inchul Song, and Yoshua Bengio.
\newblock Dynamic layer normalization for adaptive neural acoustic modeling in
  speech recognition.
\newblock {\em arXiv preprint arXiv:1707.06065}, 2017.

\bibitem{cov2016youtube}
Paul Covington, Jay Adams, and Emre Sargin.
\newblock Deep neural networks for youtube recommendations.
\newblock In {\em Proceedings of the 10th ACM Conference on Recommender
  Systems}, pages 191--198. ACM, 2016.

\bibitem{weston2013interests}
Jason Weston, Ron~J Weiss, and Hector Yee.
\newblock Nonlinear latent factorization by embedding multiple user interests.
\newblock In {\em Proceedings of the 7th ACM Conference on Recommender
  Systems}, pages 65--68. ACM, 2013.

\bibitem{ruder2016overview}
Sebastian Ruder.
\newblock An overview of gradient descent optimization algorithms.
\newblock {\em arXiv preprint arXiv:1609.04747}, 2016.

\bibitem{micikevicius2017mixed}
Paulius Micikevicius, Sharan Narang, Jonah Alben, Gregory Diamos, Erich Elsen,
  David Garcia, Boris Ginsburg, Michael Houston, Oleksii Kuchaiev, Ganesh
  Venkatesh, et~al.
\newblock Mixed precision training.
\newblock {\em arXiv preprint arXiv:1710.03740}, 2017.

\bibitem{FusionStitching}
Guoping Long, Jun Yang, Kai Zhu, and Wei Lin.
\newblock Fusionstitching: Deep fusion and code generation for tensorflow
  computations on gpus.
\newblock {\em arXiv preprint arXiv:1811.05213}, 2018.

\bibitem{jeon2018multi}
Myeongjae Jeon, Shivaram Venkataraman, Junjie Qian, Amar Phanishayee, Wencong
  Xiao, and Fan Yang.
\newblock Multi-tenant gpu clusters for deep learning workloads: Analysis and
  implications.
\newblock Technical report, MSR-TR-2018, 2018.

\bibitem{cortez2017resource}
Eli Cortez, Anand Bonde, Alexandre Muzio, Mark Russinovich, Marcus Fontoura,
  and Ricardo Bianchini.
\newblock Resource central: Understanding and predicting workloads for improved
  resource management in large cloud platforms.
\newblock In {\em Proceedings of the 26th Symposium on Operating Systems
  Principles}, pages 153--167. ACM, 2017.

\bibitem{gao2018data}
Wanling Gao, Jianfeng Zhan, Lei Wang, Chunjie Luo, Daoyi Zheng, Fei Tang, Biwei
  Xie, Chen Zheng, Xu~Wen, Xiwen He, et~al.
\newblock Data motifs: a lens towards fully understanding big data and ai
  workloads.
\newblock {\em arXiv preprint arXiv:1808.08512}, 2018.

\bibitem{guignard2018performance}
Mauricio Guignard, Marcelo Schild, Carlos~S Bederi{\'a}n, Nicol{\'a}s Wolovick,
  and Augusto~J Vega.
\newblock Performance characterization of state-of-the-art deep learning
  workloads on an ibm" minsky" platform.
\newblock In {\em Proceedings of the 51st Hawaii International Conference on
  System Sciences}, 2018.

\bibitem{chien2018characterizing}
Steven~WD Chien, Stefano Markidis, Chaitanya~Prasad Sishtla, Luis Santos, Pawel
  Herman, Sai Narasimhamurthy, and Erwin Laure.
\newblock Characterizing deep-learning i/o workloads in tensorflow.
\newblock {\em arXiv preprint arXiv:1810.03035}, 2018.

\bibitem{venkataraman2016ernest}
Shivaram Venkataraman, Zongheng Yang, Michael Franklin, Benjamin Recht, and Ion
  Stoica.
\newblock Ernest: efficient performance prediction for large-scale advanced
  analytics.
\newblock In {\em 13th $\{$USENIX$\}$ Symposium on Networked Systems Design and
  Implementation ($\{$NSDI$\}$ 16)}, pages 363--378, 2016.

\bibitem{bakhoda2009analyzing}
Ali Bakhoda, George~L Yuan, Wilson~WL Fung, Henry Wong, and Tor~M Aamodt.
\newblock Analyzing cuda workloads using a detailed gpu simulator.
\newblock In {\em 2009 IEEE International Symposium on Performance Analysis of
  Systems and Software}, pages 163--174. IEEE, 2009.

\bibitem{justus2018predicting}
Daniel Justus, John Brennan, Stephen Bonner, and Andrew~Stephen McGough.
\newblock Predicting the computational cost of deep learning models.
\newblock In {\em 2018 IEEE International Conference on Big Data (Big Data)},
  pages 3873--3882. IEEE, 2018.

\end{thebibliography}

\end{document}